\documentclass[aps,twocolumn,groupedaddress]{revtex4}
\usepackage{epsfig}
\usepackage{graphicx}
\usepackage[T1]{fontenc}
\usepackage{ae}
\usepackage{color}
\usepackage[latin1]{inputenc}
\newcommand {\mofe} {$\textrm{Mo}_{72}\textrm{Fe}_{30}$}
\newcommand {\mocr} {$\textrm{Mo}_{72}\textrm{Cr}_{30}$}
\newcommand {\mov} {$\textrm{Mo}_{72}\textrm{V}_{30}$}
\newcommand {\wv} {$\textrm{W}_{72}\textrm{V}_{30}$}

\begin{document}

%\begin{center}\today\end{center}

%%%%%%%%%%%%%%%%%%%%%%%%%%%%%%%%%%%%%%%%%%%%%%%%%%%%%%%%%%%%%%%%%%%%%%%%%%%%%

\title[HTE of Heisenberg Hamiltonians ]
{Eighth-order high-temperature expansion for general Heisenberg
Hamiltonians}

\author{Heinz-J\"urgen Schmidt$^1$
\footnote[3]{Correspondence should be addressed to
hschmidt@uos.de}
, Andre Lohmann$^2$ and Johannes Richter$^2$ }

\address{$^1$Universit\"at Osnabr\"uck, Fachbereich Physik,
Barbarastr. 7, D - 49069 Osnabr\"uck, Germany\\
$^2$Institut f\"ur Theoretische Physik, Otto-von-Guericke-Universit\"at Magdeburg,\\
PF 4120, D - 39016 Magdeburg, Germany }

%\tableofcontents

\begin{abstract}
We explicitly calculate the moments $t_n$ of general Heisenberg
Hamiltonians up to eighth order. They have the form of finite sums
of products of two factors. The first factor is represented by a
(multi-)graph which has to be evaluated for each particular system
under consideration. The second factors are well-known universal
polynomials in the variable $s(s+1)$, where $s$ denotes the
individual spin quantum number. From these moments we determine the
corresponding coefficients of the high-temperature expansion of the
free energy and the zero field susceptibility by a new method. These
coefficients can be written in a form which makes explicit their
extensive character. Our results represent a general tool to
calculate eighth-order high-temperature series for arbitrary
Heisenberg models. The results are applied to concrete systems,
namely to magnetic molecules with the geometry of the
icosidodecahedron, to frustrated square lattices, and to the
pyrochlore magnets. By comparison with other methods that have been
recently applied to these systems, we find that the typical
susceptibility maximum of the spin-$s$ Heisenberg antiferromagnet is
well described by the eighth-order high-temperature series.
\end{abstract}

\maketitle

\section{Introduction}\label{sec:I}
The Heisenberg model
\begin{equation}\label{model}
H=\sum_{\mu <\nu} J_{\mu \nu} {\bf s}_\mu \cdot {\bf s}_\nu
\end{equation}
is the basic model to describe physical properties of magnetic
insulators. Despite its simplicity the thermodynamics of the model
is generally unknown. For unfrustrated quantum spin systems the
quantum Monte Carlo (QMC) method provides accurate numerical results
for the temperature dependence of the physical quantities. If the
exchange couplings are frustrated the so-called ``sign problem"
precludes accurate QMC calculations\cite{TrWi05}. For
one-dimensional (1D) frustrated systems the density-matrix
renormalization group approach\cite{schollwoeck} yields precise
results in the whole temperature range. For frustrated quantum spin
systems in dimension $D>1$ accurate methods to calculate
thermodynamic properties are notoriously rare. Quite reasonable
results for arbitrary temperatures $T$ can be obtained, e.g., by a
second-order Green function technique, see, e.~g.~,
Ref.~\onlinecite{froebrich,rgm1,rgm2}. However, the application of
this method needs quite a lot of technical experience. Hence, a
simple but universal approach is desirable. A well established
method fulfilling this criterion is the high-temperature expansion
(HTE).  Since often experimental results, e.~g.~, for the
susceptibility, are available in a wide temperature range (including
temperatures exceeding the energy scale set by the major exchange
constant $J$, i.~e.~, for $k T\gg |J|$), the HTE can serve as a
method to extract the exchange constants of the Heisenberg model
from experimental data.

For Heisenberg models on the simple two-dimensional (2D) and
three-dimensional (3D) lattices the HTE is available up to high
orders, see Refs.~\onlinecite{domb_green} and \onlinecite{OHZ06} and references
therein. However, often one is faced with materials where two or
even more different exchange constants are relevant. A typical
example are frustrated quasi-1D or quasi-2D magnets where except the
nearest-neighbor (NN) and next-nearest-neighbor (NNN) in-chain or
in-plane couplings also the interchain  or interplane  couplings
are important.
Typically, for such more complex exchange geometries
the HTE is known only up to low order. In this situation it would be
desirable to have at one's disposal explicit formulas of higher
order HTE for general Heisenberg systems and general spin quantum
number $s$.  It is the aim of the present paper to derive such
formulas. The key notion is given by the the moments $\mbox{Tr }H^n$
of order $n$, which can be expressed as sums over suitable sets of
graphs. From the moments one can derive the coefficients of the HTE
for, say, susceptibility or specific heat in a tedious but straightforward manner.
Unfortunately, the number of involved graphs grows
super-exponentially with the order $n$, which delimits the maximal
order of the HTE for practical purposes. In this paper, we have
confined ourselves to calculations up to eighth order and have to
take account of $1139$ relevant graphs. Nevertheless, this order is
sufficient to describe typical properties of frustrated spin
systems, as we will show by means of examples.

The calculation of the HTE for spin systems has a long tradition.
Since the $1970$s it is known that the moments of certain spin lattices with
only one exchange constant can be written as sums over sets of graphs
${\mathcal G}_\nu$ with two factors.
The first factor was called the ``lattice constant" and counts how often the
graph ${\mathcal G}_\nu$ can be embedded into the spin lattice.
The second factor is a universal polynomial $p_\nu(r)$ in the variable $r=s(s+1)$.
The polynomials $p_\nu(r)$ up to eighth order together with the corresponding graphs
${\mathcal G}_\nu$ are contained in the appendix of Ref.~\onlinecite{domb_green}.
We have independently calculated these polynomials by computer-algebraic means and
confirmed a sample of the data in Ref.~\onlinecite{domb_green}.
The generalization of these results from simple spin lattices to arbitrary Heisenberg models
is achieved by replacing the above-mentioned ``lattice constant" by an ``evaluation" of the graph
${\mathcal G}_\nu$ for the spin system under consideration. This evaluation involves sums of
products of coupling constants $J_{\mu\nu}$ and yields analytical expressions for the moments
of $H$ and the coefficients of the HTE of susceptibility and specific heat.
It seems that such general analytical expressions for moments have
only be published up to order three, see Ref.~\cite{SSL01}. Those
papers that consider higher-order expansions, see,
e.~g.~, Refs.~\cite{wood1957,pirnie1966,TWB10,CGSM08,FHW04,OZ04a,OZ04b,LKMW03,ST02,HL01,BEU00,ZHO99,ES98a,ES98b,OB96,KBJ96,Jetal00,Rosner_HTE,MBP03}
are usually confined to special cases, i.~e.~, special geometries or special values of $s$.
We have used some of these papers, namely
Refs.~\onlinecite{wood1957,pirnie1966} and \onlinecite{Rosner_HTE}
to check our general results.

The paper is organized as follows. In Sec.~\ref{sec:D} we give the definitions used
and illustrate the underlying mathematics. In Sec.~\ref{sec:R}
we present general results of the HTE coefficients up to fourth order for the
moments of the Hamiltonian, the free energy, the specific heat,  the
magnetic moments, and the susceptibility.
The very general expressions up to eighth order can be
found in Supplementary Material 1\cite{supp1} of Ref.~\cite{SLR11}.
In Sec.~\ref{applic}
we apply our method to specific Heisenberg models, which are currently
discussed in the literature, namely the Heisenberg antiferromagnet on
the Archimedean icosidodecahedron,  frustrated square-lattice
Heisenberg model as well as the Heisenberg model on the pyrochlore
lattice.
For these models the HTE for the specific heat and the susceptibility up to eighth order
for arbitrary spin quantum number $s$ are collected in the appendices
and  Supplementary Material 2\cite{supp2} of Ref.~\cite{SLR11}.

Although, the information provided in this paper and the
supplementary materials\cite{supp1} allows, in principle, to calculate
the HTE up to eighth order, it might be a tedious task to do so in
practice. Hence, we provide a simple computer program written in
 C++ that allows to calculate within  a few seconds the eighth-order HTE
coefficients as well as the Pad\'e approximants for the
susceptibility and the specific heat for an arbitrary Heisenberg
model with up to four different exchange constants\cite{program}.

\section{Definitions}\label{sec:D}
In this paper we consider systems of $N$ spins with individual spin
quantum number $s=\frac{1}{2},1,\frac{3}{2},\ldots$. The Heisenberg
Hamiltonian has the form (\ref{model}) where the
$J_{\mu\nu}=J_{\nu\mu},\,1\le\mu\neq\nu\le N$ are suitable coupling
constants and $\mathbf{s}_\mu$ denotes the spin vector operator of
the $\mu$th spin. The moments $t_n$ of $H$ will be normalized by
division by the dimension of the total Hilbert space,
i.~e.~, $t_n\equiv\frac{\mbox{\scriptsize Tr}(H^n)}{(2s+1)^N}$.
Analogously, the magnetic moments of $H$ are defined by
$\mu_n=\frac{\mbox{\scriptsize
Tr}(S^{(3)2}H^n)}{(2s+1)^N}$, where $\mathbf{S}$ denotes
the total spin vector and ${S}^{(i)},\;\;i=1,2,3,$ its $i$th
component. As usual, $\chi(\beta)=\beta \frac{\mbox{\scriptsize
Tr}({S}^{(3)2}\exp(-\beta H))} {\mbox{\scriptsize
Tr}(\exp(-\beta H))}$ denotes the normalized zero field
susceptibility. $\chi(\beta)=\sum_{n=1}^{\infty} c_n \beta^n$ is its
HTE in terms of the dimensionless
inverse temperature $\beta\equiv\frac{|J|}{k\,T}$, where $J$ is a
typical energy. The Hamiltonian $H$ is understood to be
dimensionless upon division by $|J|$. The free energy $F(\beta)$ is
defined by $-\beta F(\beta)=\ln\left( \mbox{Tr } e^{-\beta
H}\right)$ and its HTE is given by $-\beta
F(\beta)=\sum_{n=0}^{\infty} a_n \beta^n$. From this one derives the
normalized specific heat $C(\beta)\equiv -2\beta^2\frac{\partial
F}{\partial \beta}-\beta^3\frac{\partial^2 F}{\partial \beta^2}$ and
a short calculation shows that its HTE $C(\beta)=\sum_{n=2}^{\infty}
d_n \beta^n$ is related to that of $F(\beta)$ by $d_n=n(n-1)a_n$ for
$n=2,3,\ldots$. \\

HTE are usually written in a compact way by utilizing
graph-theoretic notations, see, e.~g.~, Refs.~\onlinecite{domb_green}
and \onlinecite{OHZ06}.
Let ${\mathcal G}$
be a multigraph consisting of $g$ nodes (vertices) and
a number of ${\mathcal N}(i,j)={\mathcal N}(j,i)$ bonds (edges) between the
$i$th and the $j$th node. We do not consider ``loops,"
i.~e.~, ${\mathcal N}(i,i)=0$ for all $i=1,\ldots,g$. The total number
of all bonds, $\gamma({\mathcal G})=\sum_{i<j}{\mathcal N}(i,j)$
will be called the size of ${\mathcal G}$. ${\mathcal
G}$ is not necessarily connected, see the examples below. We will
identify the set of $g$ nodes with $\{1,2,\ldots,g\}$ and the set of
$N$ spins with $\{1,2,\ldots,N\}$. To simplify the wording we will
omit the prefix ``multi-" and simply speak of ``graphs"
in what follows. A selection of graphs ${\mathcal
G}_\nu,\;\nu=1,\ldots,$ needed for purposes of illustration is
represented in table \ref{table1}.
A complete list of all relevant graphs up to size $8$ can be found in
Supplement 1.\cite{supp1}\\

For every graph we define its multinomial factor by
\begin{equation}\label{dfac}
f({\mathcal G})\equiv \frac{\gamma({\mathcal G})!}
{\prod_{i<j}{\mathcal N}(i,j)!} \;.
\end{equation}

Define the symmetry group $G({\mathcal G})$ of a graph
in the obvious way
\begin{eqnarray}\nonumber
G({\mathcal G})&\equiv& \{\pi\in{\mathcal S}_g | {\mathcal
N}(i,j)={\mathcal N}(\pi(i),\pi(j))\\ \label{dsym}
&& \mbox{ for all } 1\le i,j \le g
\} \;.
\end{eqnarray}
Here ${\mathcal S}_g$
denotes the group of all permutations
$\pi:\{1,\ldots,g\}\longrightarrow\{1,\ldots,g\}$.
A localization of a graph ${\mathcal G}$ is an embedding
\begin{equation}\label{demb}
\jmath:\{1,\ldots,g\}\longrightarrow \{1,\ldots,N\}
\end{equation}
up to symmetries of ${\mathcal G}$. More precisely, two embeddings
$\jmath_1,\jmath_2:\{1,\ldots,g\}\longrightarrow \{1,\ldots,N\}$ are
called equivalent if and only if $\jmath_1=\jmath_2\circ\pi$ for some $\pi\in
G({\mathcal G})$, and a localization of ${\mathcal G}$ is a
corresponding equivalence class of embeddings. The number of
localizations of ${\mathcal G}$ (for given $N$) will be denoted by
$L$. We will also speak of localized graphs ${\mathcal G}$
which will be represented by attaching numbers of different spin
sites to the nodes of ${\mathcal G}$, with the understanding that
two localized graphs which only differ by a symmetry permutation of
the spin sites are considered as identical,
e.~g.~,
$\begin{array}{c}{\includegraphics[width=1cm]{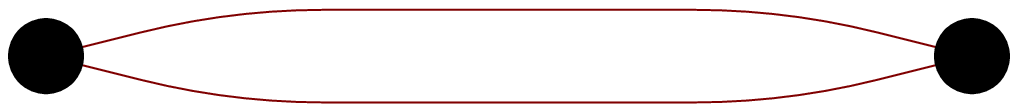}}\\
1\quad\quad 2
\end{array}=\begin{array}{c}{\includegraphics[width=1cm]{G2}}\\
2\quad\quad 1
\end{array}$.\\

Two localized graphs ${\mathcal G}_1,{\mathcal G}_2$ can be
soldered in a natural way yielding the ``soldering
product" ${\mathcal G}_1\oplus {\mathcal G}_2$, which is another
localized graph. The nodes of ${\mathcal G}_1\oplus {\mathcal G}_2$
are identified according to their numbering and the bonds are
correspondingly added. For example,
\begin{equation}\label{dec}
{\includegraphics[width=10mm]{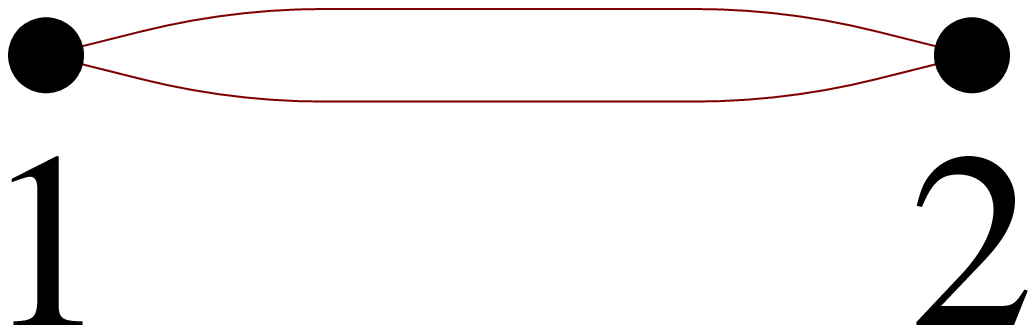}} \oplus
{\includegraphics[width=20mm]{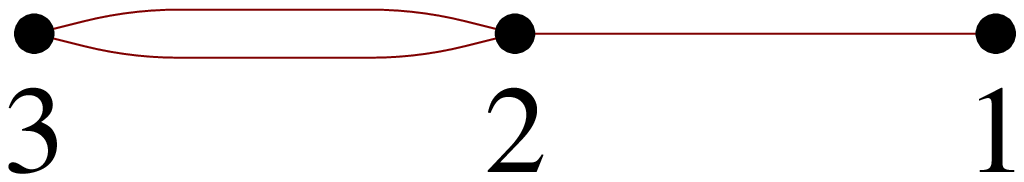}}
={\includegraphics[width=20mm]{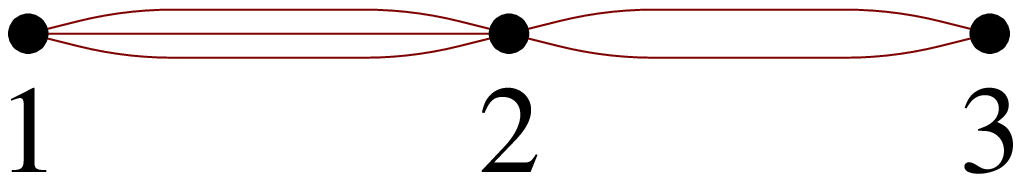}}
\end{equation}
\\

Conversely, we will say that the localized graph ${\mathcal
G}_1\oplus {\mathcal G}_2$ is decomposed into ${\mathcal
G}_1$ and ${\mathcal G}_2$.
In general, a localized graph can be decomposed into different ways.\\

From the expansion
\begin{equation}\label{dexp}
\mbox{Tr}H^n=\sum_{\mu_1<\nu_1,\ldots,\mu_n<\nu_n}\;\prod_{i}J_{\mu_i
\nu_i}\;\mbox{ Tr }
\left(\prod_{i}\bf{s}_{\mu_i}\cdot\bf{s}_{\nu_i}\right)
\end{equation}
it is clear that the expressions for the moments $t_n$ involve
various products of coupling constants $J_{\mu\nu}$. The structure
of these products can be represented by the graphs ${\mathcal G}$
defined above, such that the factors $J_{\mu\nu}^\ell$ correspond to
the bonds of ${\mathcal G}$ with multiplicity $\ell$. The sum of
different products in (\ref{dexp}) of the same structure will be
obtained by an \underline{evaluation} of ${\mathcal G}$, denoted by
$\overline{\mathcal G}$, for the spin system under consideration.
$\overline{\mathcal G}$ denotes a real number which depends on the
coupling constants and only implicitly on the number $N$ of spins.
This number will be defined according to the following statements:
\begin{enumerate}
\item If $g>N$ we set $\overline{\mathcal G}=0$.\\
\item If $g\le N$ we select from each equivalence class of embeddings
a certain representative
\begin{equation}\label{dembl}
\jmath_\ell:\{1,\ldots,g\}\longrightarrow
\{1,\ldots,N\},\;\ell=1,\ldots,L
\end{equation}
and define
\begin{equation}\label{d2}
\overline{\mathcal G}\equiv \sum_{\ell=1}^L\, \prod_{1\le i<j\le g}
\, \left( J_{\jmath_\ell(i),\jmath_\ell(j)} \right)^{{\mathcal
N}(i,j)} \;.
\end{equation}
\end{enumerate}
Obviously, the definition of $\overline{\mathcal G}$ does not depend
on the choice of representatives $\jmath_\ell$ since the product
$\prod_{1\le i<j\le g}
\,\left(J_{\jmath_\ell(i),\jmath_\ell(j)}\right)^{{\mathcal
N}(i,j)}$ is invariant under
permutations from the symmetry group $\pi\in G({\mathcal G})$.\\

In order to illustrate this definition we consider an example of
$N=4$ spins and ${\mathcal G}={\includegraphics[width=5mm]{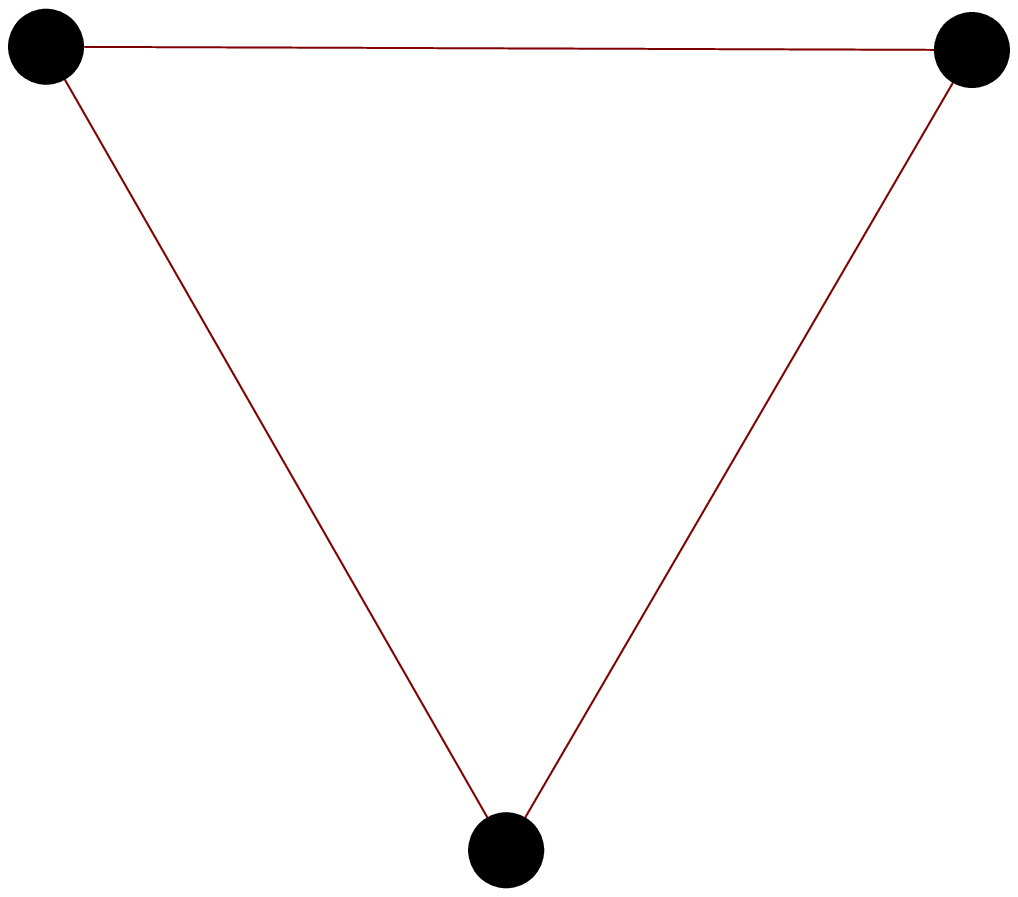}}$,
hence $g=3<4=N$. The symmetry group $G({\mathcal G})$ consists of
all permutations of $\{1,2,3\}$ hence $|G({\mathcal G})|=3!=6$.
There are $4!$ embeddings
$\jmath:\{1,2,3\}\longrightarrow\{1,2,3,4\}$ and $L=\frac{4!}{3!}=4$
equivalence classes from which we choose the representatives
\begin{eqnarray}\label{dj1234}
\jmath_1&=&(1\rightarrow 1,\,2\rightarrow 2,\,3\rightarrow 3),\\
\jmath_2&=&(1\rightarrow 1,\,2\rightarrow 2,\,3\rightarrow 4),\\
\jmath_3&=&(1\rightarrow 1,\,2\rightarrow 3,\,3\rightarrow 4),\\
\jmath_4&=&(1\rightarrow 2,\,2\rightarrow 3,\,3\rightarrow 4)\;.
\end{eqnarray}
Hence $\overline{{\mathcal G}}
=J_{12}J_{23}J_{13}+J_{23}J_{34}J_{23}+J_{34}J_{14}J_{13}+J_{14}J_{12}J_{24}$.\\

The coefficients $c_n$ of the susceptibility's HTE (and similarly
the $a_n$ of the free energy HTE) will contain products of
evaluations $\overline{{\mathcal G}_\nu}\;\overline{{\mathcal
G}_\mu}$. These expressions can be simplified using rules which
transform such products into linear combinations of other
evaluations. To give an example, we consider
$\overline{{\mathcal G}_1}\;\overline{{\mathcal G}_2}=
\overline{\makebox[5mm]{\rule{0mm}{2.5mm}\includegraphics[width=5mm]{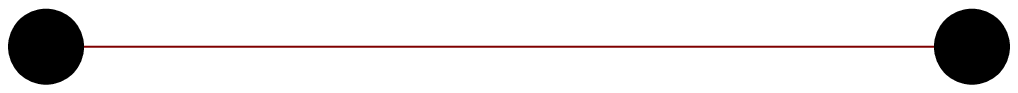}}}\;
\overline{\makebox[5mm]{\rule{0mm}{2.5mm}\includegraphics[width=5mm]{G2}}}=
\left(\sum_{\mu<\nu}J_{\mu\nu}\right)\,\left(\sum_{\kappa<\lambda}J_{\kappa\lambda}^2\right)$.
It is obvious that this product can be written as a sum over
evaluations of the three graphs which can be combined from
${\includegraphics[width=10mm]{G1}}$ and
${\includegraphics[width=10mm]{G2}}$, namely
${\includegraphics[width=10mm]{G1}}{\includegraphics[width=10mm]{G2}}$,
${\includegraphics[width=20mm]{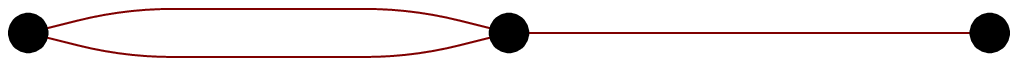}}$ and
${\includegraphics[width=10mm]{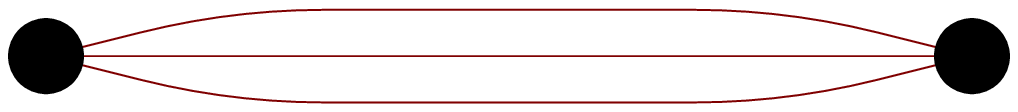}}$. In fact,
\begin{equation}
\overline{\makebox[10mm]{\rule{0mm}{2.5mm}\includegraphics[width=10mm]{G1}}}\;\;
\overline{\makebox[10mm]{\rule{0mm}{2.5mm}\includegraphics[width=10mm]{G2}}}\;
=\;\overline{\makebox[10mm]{\rule{0mm}{2.5mm}\includegraphics[width=10mm]{G1}}
\makebox[10mm]{\rule{0mm}{2.5mm}\includegraphics[width=10mm]{G2}}}\;
+\;\overline{\makebox[20mm]{\rule{0mm}{2.5mm}\includegraphics[width=20mm]{G6}}}\;
+\;\overline{\makebox[10mm]{\rule{0mm}{2.5mm}\includegraphics[width=10mm]{G5}}}\;.
\end{equation}
\\

Similar expressions can be derived for other products of evaluations
yielding various ``product rules" of the form
\begin{equation}\label{drules}
\overline{{\mathcal G}_\mu}\;\overline{{\mathcal G}_\nu} =
\sum_{\lambda}c_{\mu\nu}^\lambda \;\overline{{\mathcal
G}}_\lambda\;.
\end{equation}
Here the sum over $\lambda$ runs through all graphs ${\mathcal
G}_\lambda$ whose localizations are soldering products of
localizations of ${\mathcal G}_\mu$ and ${\mathcal G}_\nu$. The
integers $c_{\mu\nu}^\lambda$ count the number of ways to decompose
a localization of ${\mathcal G}_\lambda$ into localizations of
${\mathcal G}_\mu$ and ${\mathcal G}_\nu$. For example, the
decomposition (\ref{dec}) is unique (up to symmetries), hence
$c_{2,6}^{35}=1$, c.~f.~table \ref{table1}. On the other hand,
\begin{eqnarray}
{\includegraphics[width=20mm]{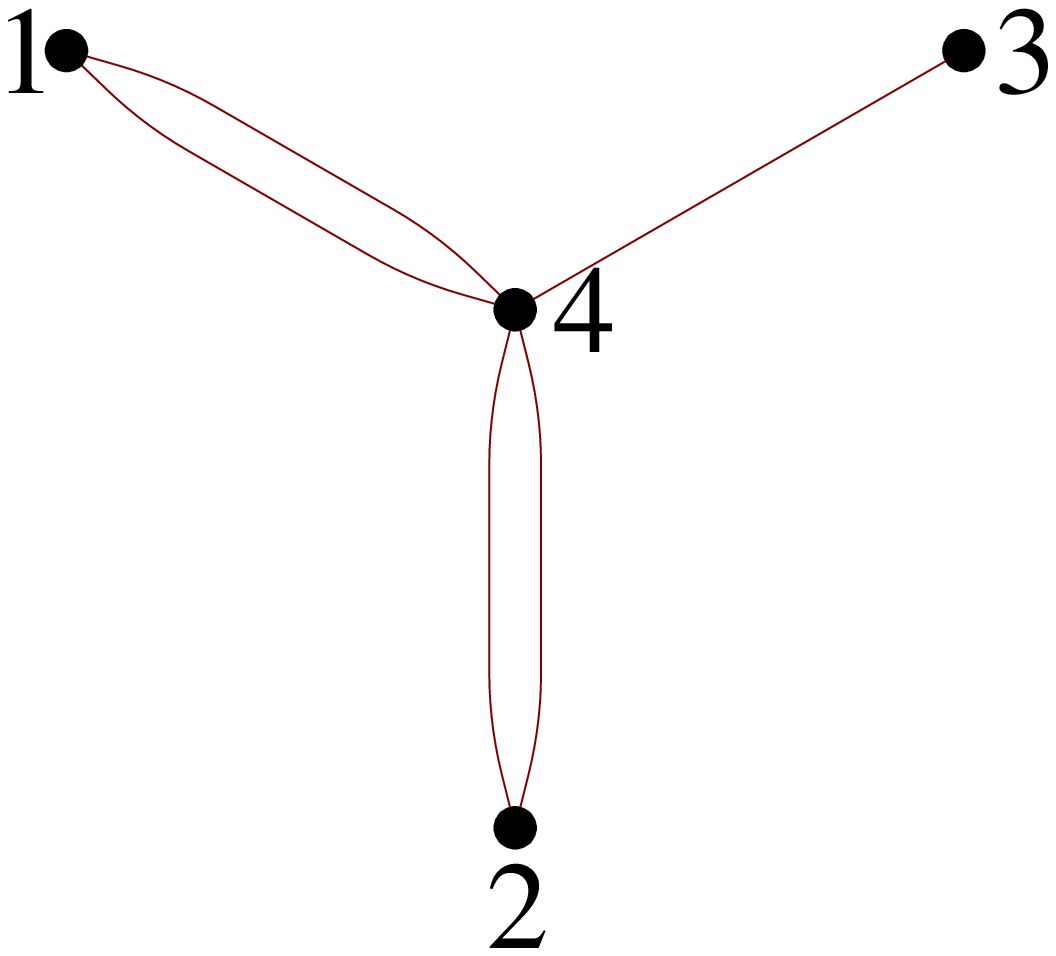}}&=&{\includegraphics[width=10mm]{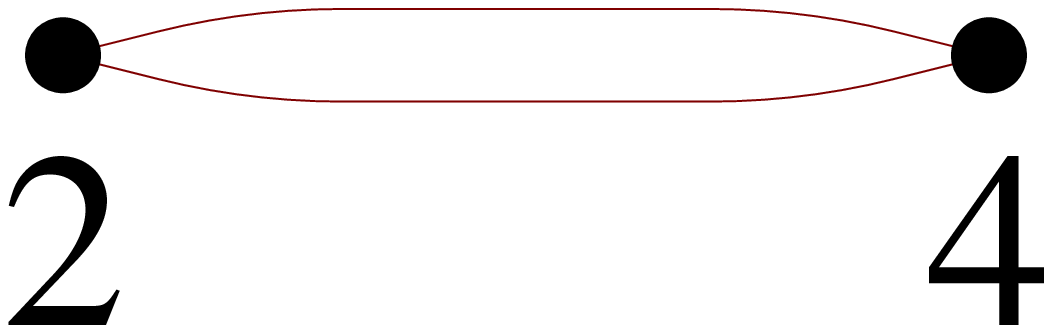}}\oplus
{\includegraphics[width=20mm]{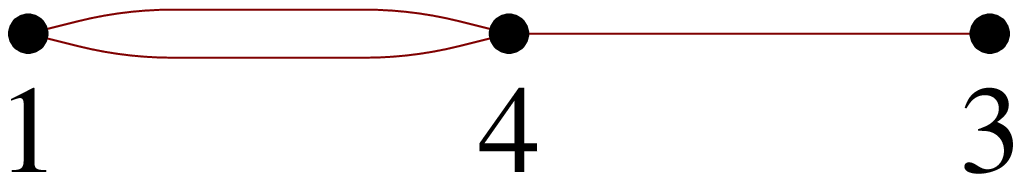}}\\
&=&{\includegraphics[width=10mm]{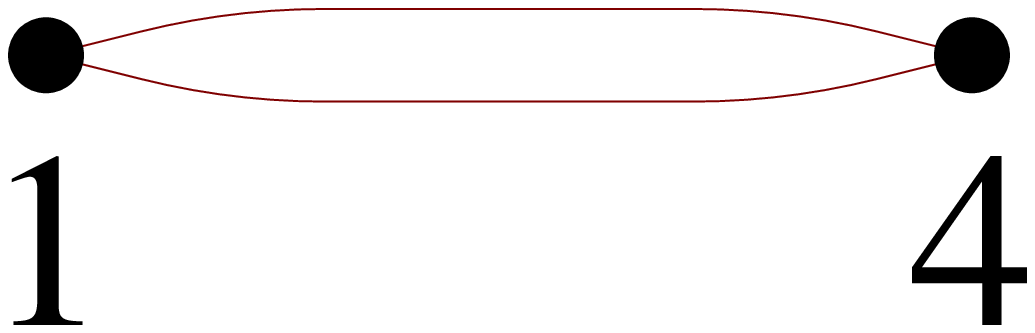}}\oplus
{\includegraphics[width=20mm]{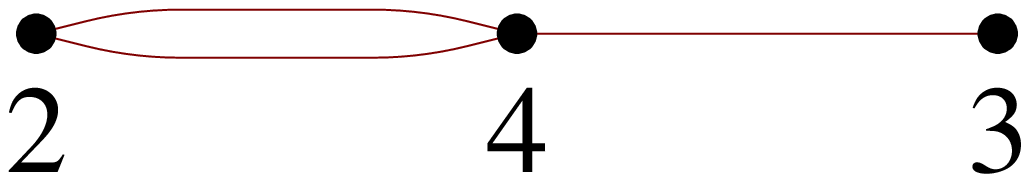}},
\end{eqnarray}
hence
$c_{2,6}^{39}=2$, c.~f.~table \ref{table1}.\\
In the case ${\mathcal G}_\mu={\mathcal G}_\nu$ we have to define
$c_{\mu\nu}^\lambda$ in such a way that the binomial factor $2$ is
included for products of different localizations. For example,
\begin{equation}
\overline{\makebox[5mm]{\rule{0mm}{2.5mm}\includegraphics[width=5mm]{G2}}}^2
=\;\overline{\makebox[10mm]{\rule{0mm}{2.5mm}\includegraphics[width=10mm]{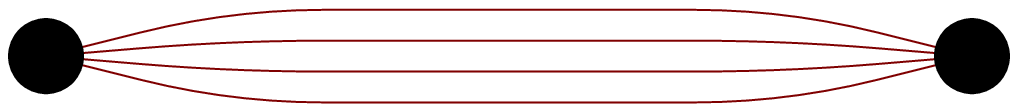}}}\;
+2\;\overline{\makebox[10mm]{\rule{0mm}{2.5mm}\includegraphics[width=10mm]{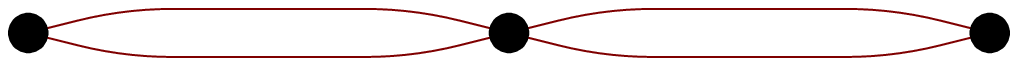}}}\;
+2\overline{\includegraphics[width=10mm]{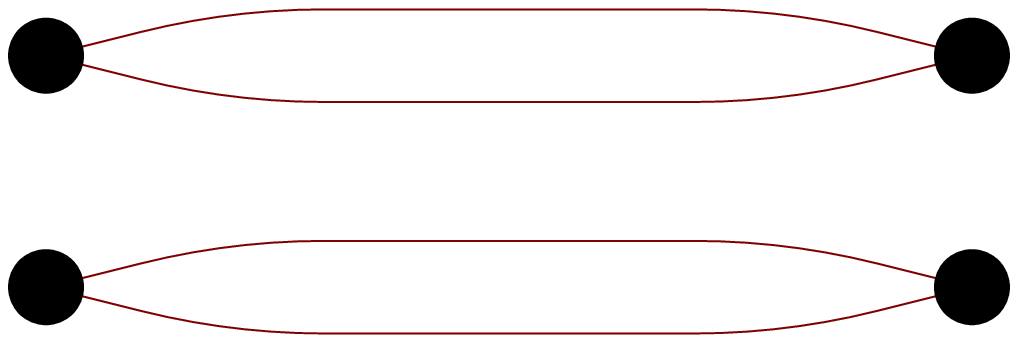}}.
\end{equation}
\\

From the product rules (\ref{drules}) one can derive further ones
for multiple products.

\begin{table}[ht]
\caption{A selection of graphs ${\mathcal G}_\nu$ \label{table1}}
\begin{center}
\begin{tabular}{||l|l||l|l||l|l||}
\hline \\
{$\nu$ }& ${\mathcal G}_\nu$ &$\ldots$ & $\ldots$ &$\ldots$ & $\ldots$ \\
\hline
\hline\\
 {$1$}  &{\includegraphics[width=2cm]{G1}} &$2$  &\includegraphics[width=2cm]{G2}
 &$3$  &\includegraphics[width=2cm]{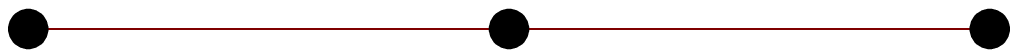} \\
\hline\\
$5$  &\includegraphics[width=2cm]{G5} &$6$
&\includegraphics[width=2cm]{G6}
 &$8$  &\includegraphics[width=2cm]{G8}  \\
\hline\\
$9$  &\includegraphics[width=2cm]{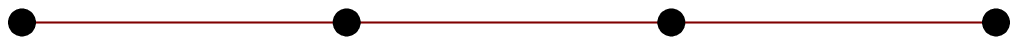} &$10$
&\includegraphics[width=2cm]{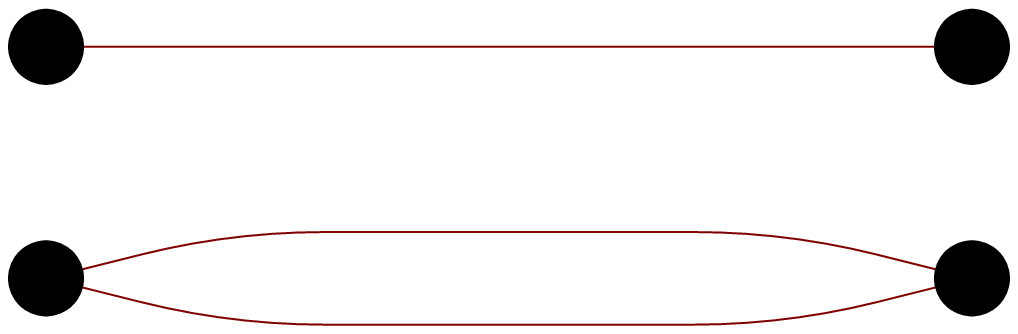}
 &$12$  &\includegraphics[width=2cm]{G12}   \\
\hline\\
$14$  &\includegraphics[width=2cm]{G14}&$16$
&\includegraphics[width=2cm]{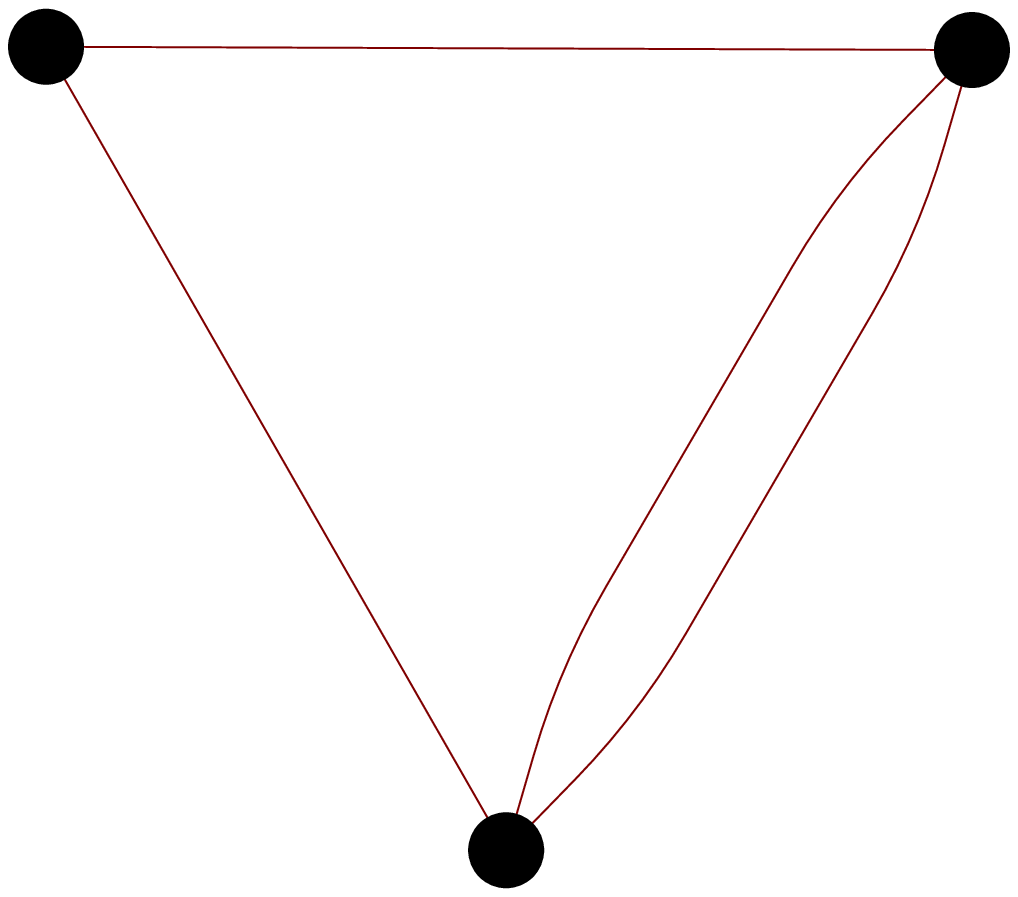}
 &$23$  &\includegraphics[width=2cm]{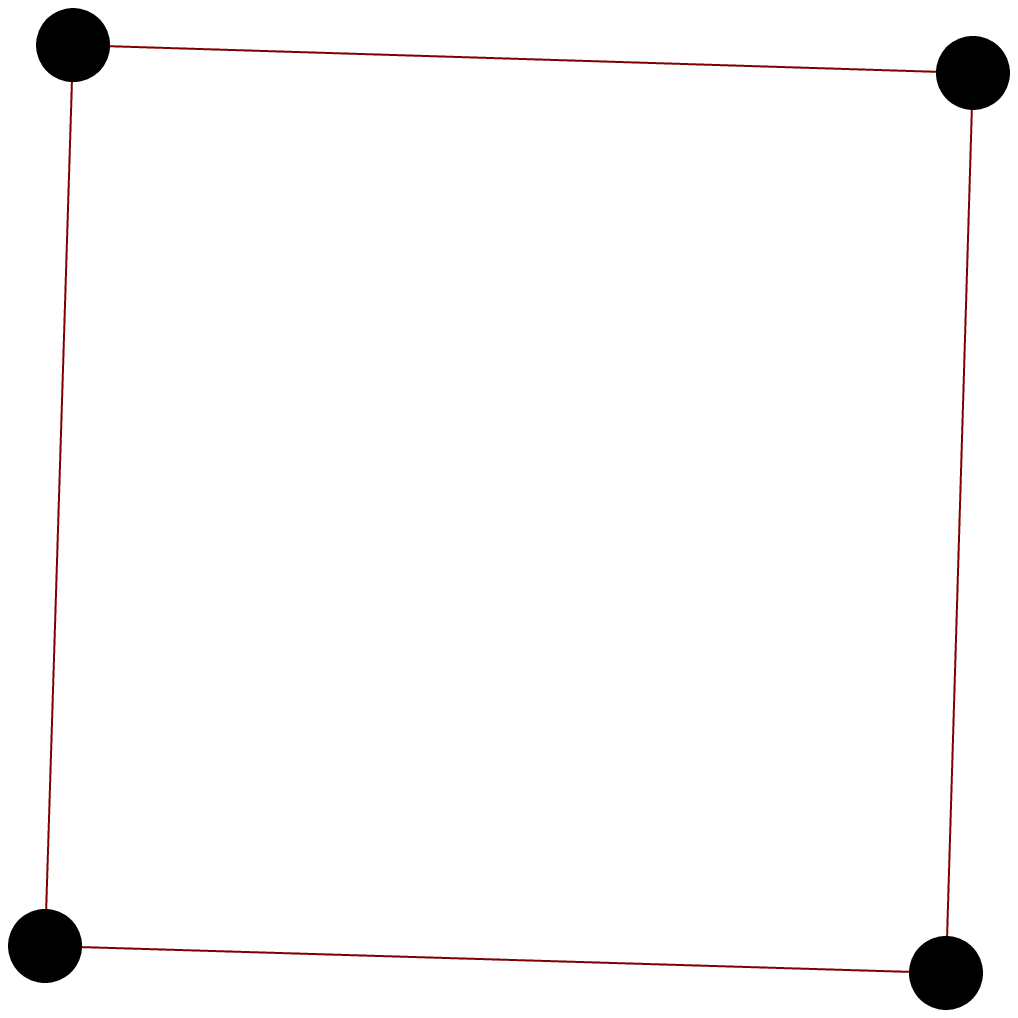}    \\
\hline\\
$27$  &\includegraphics[width=2cm]{G27}& $35$
&\includegraphics[width=2cm]{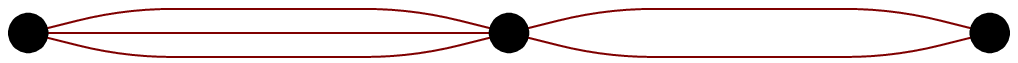}
 &  $39$  &\includegraphics[width=2cm]{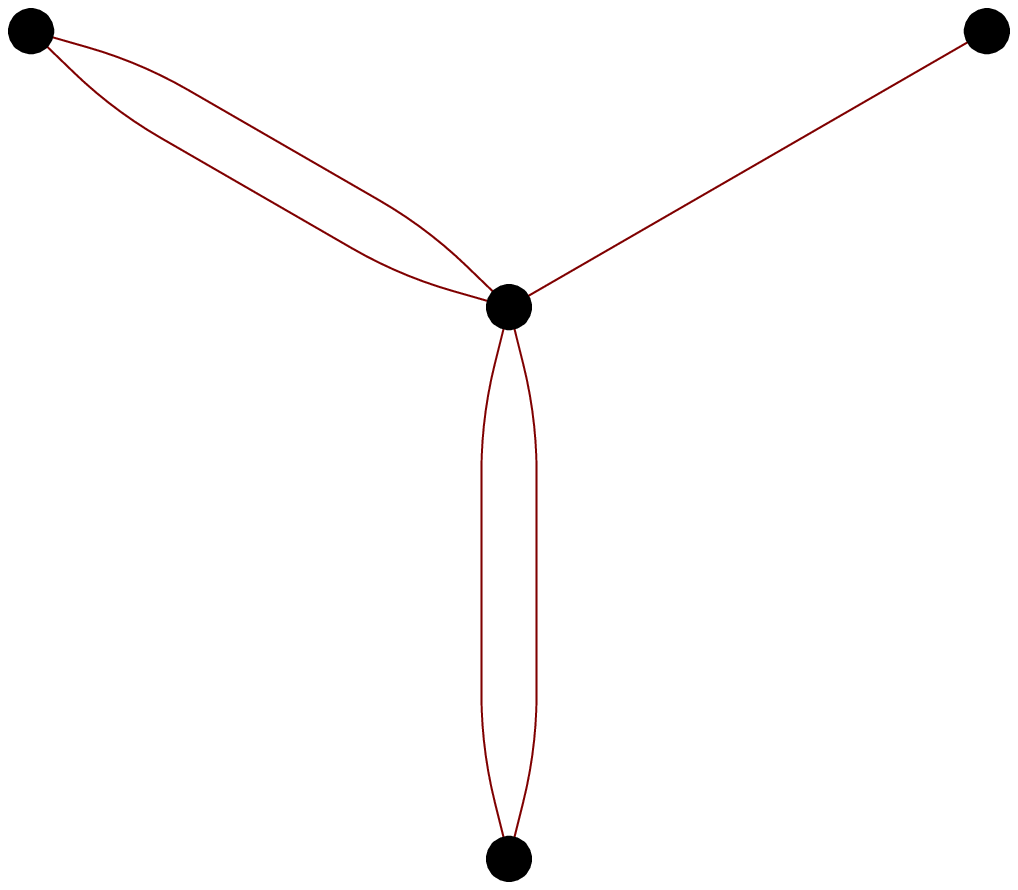}   \\
\hline
\end{tabular}
\end{center}
\end{table}

\section{Results}\label{sec:R}
\subsection{Moments}\label{sec:R_mom}
It turns out that the moments $t_n$ can be written in the following
way:
\begin{equation}\label{r1}
t_n=\sum_{\nu\in T_n}\;\overline{\mathcal G}_\nu\; p_\nu(r) \;.
\end{equation}
Here the ${\mathcal G}_\nu,\,\nu\in T_n,$ denote certain graphs of
size $n$ and the $p_\nu$ are polynomials of order $\le n$ in the
variable $r=s(s+1)$. Actually, the $p_\nu$ are of the form
$p_\nu=\sum_{i=g}^n a^{(\nu)}_i\,r^i$ where $g$ denotes the order of
${\mathcal G}_\nu$ and some $a^{(\nu)}_i$ may vanish. The leading
coefficients $a^{(\nu)}_n$ determine the classical limit
$r\longrightarrow \infty$ of the moments, hence they can be
calculated by means of integrals
over unit spheres.\\

It is crucial that the polynomials $p_\nu$ depend neither on $N$ nor
on the coupling constants $J_{\mu\nu}$ whereas the terms
$\overline{{\mathcal G}_\nu}$ depend only on the coupling constants
and only implicitly on $N$ via (\ref{d2}). The polynomials $p_\nu$
up to eighth order are well known and have been used for the HTE of
certain spin lattices.  A subset of the $p_\nu$ is, for example,
listed in Ref.~\cite{domb_green} together with certain rules which permit
the calculation of the remaining polynomials. The most important
rule holds in the case where ${\mathcal G}$ is the disjoint union of
two simpler graphs, ${\mathcal G}={\mathcal G}_1 \biguplus {\mathcal
G}_2$ and reads
\begin{equation}\label{runion}
p({\mathcal G})=p({\mathcal G}_1)\,p({\mathcal G}_2)
\frac{f({\mathcal G})}{f({\mathcal G}_1)\,f({\mathcal G}_2)}
\;.
\end{equation}
Note that the polynomials in Ref.~\cite{domb_green} are defined as our
$p_\nu$ divided by the multinomial factor (\ref{dfac}), hence these
factors do not occur in the rule analogous to (\ref{runion}). Other
rules, which we need not repeat here, say that the $p_\nu$ vanish
{\it a priori} for certain graphs.\\

For the determination of $t_n$ it thus suffices to enumerate the
graphs ${\mathcal G}_\nu, \;\nu\in T_n$ and the corresponding
polynomials $p_\nu$. We will give the first four moments for the
sake of illustration and defer the lengthy expressions for
$t_n,\,n=5,6,7,8$ to the Supplemental Material 1.\cite{supp1}
\begin{eqnarray}\label{rta}
t_1&=&0\\ \label{rtb} t_2&=&
\sum_{\mu<\nu}J_{\mu\nu}^2\;\frac{1}{3}r^2=
\frac{1}{3}r^2\,\overline{{\mathcal G}_2},\\  \label{rtc} t_3&=&
-\frac{1}{6} r^2\; \overline{{\mathcal G}_5} + \frac{2}{3} r^3
\;\overline{{\mathcal G}_8},\\ \nonumber
 t_4&=& \frac{1}{15}r^2 (2
-2 r + 3 r^2)\; \overline{{\mathcal G}_{12}} +\frac{2}{9}r^3(-1 + 3
r)\; \overline{{\mathcal G}_{14}}\\ \label{rtd} && -\frac{2}{9}
r^3\;\overline{{\mathcal G}_{16}} +\frac{8}{9} r^4\;\overline{{\mathcal G}_{23}}
+\frac{2}{3} r^4\;\overline{{\mathcal G}_{27}}\;.
\end{eqnarray}

\subsection{Free energy}\label{sec:R_f}
It is well known that the coefficients of the power series for the
free energy $F(\beta)$
\begin{equation}\label{rf1}
-\beta F(\beta)=\ln\left( \mbox{Tr } e^{-\beta
H}\right)=\sum_{n=0}^{\infty} a_n \beta^n
\end{equation}
can be expressed in terms of the moments $t_n$ and its products. As
indicated in Sec.~\ref{sec:D}, a variety of product rules can be
used to simplify the resulting expressions. This simplification,
which is sometimes also referred to as the ``cumulant expansion",
see, e.~g.~, Ref.~\onlinecite{OHZ06}, has the further advantage that it reveals the
extensive character of the $a_n$. By this we mean the following. If
the spin system under consideration would have a periodic lattice
structure of, say, $K$  unit cells with periodic boundary
conditions, it follows immediately that the evaluation of a single
graph $\overline{{\mathcal G}}$ linearly scales with $K$, and hence
with $N$, as long as ${\mathcal G}$ is connected. For unconnected
${\mathcal G}$ the evaluation scales with $K^c$ where $c$ is the
number of connected components of ${\mathcal G}$. Obviously,
products of evaluations of connected graphs $\overline{{\mathcal
G}_\nu}\;\overline{{\mathcal G}_\mu}$ would scale with $K^2$. It
turns out that the elimination of these and higher products in the
expression for the $a_n$ by means of the rules (\ref{drules}) also
eliminates the evaluation terms of unconnected graphs. This has to
be expected on physical grounds, since the total free energy of a
spin lattice should be an extensive quantity, i.~e.~, linearly scale
with $K$.
But it is an additional consistency test of our results that the
non-extensive contributions to the $a_n$ actually cancel.\\

The first five coefficients of the series (\ref{rf1}) read as
follows:

\begin{eqnarray} \label{rf2a}
a_0&=&N\,\ln(2s+1),\\ \label{rf2b} a_1&=&0, \\ \label{rf2c}
a_2&=&\frac{1}{6}r^2\; \overline{{\mathcal G}_2},\\ \label{rf2d}
a_3&=&\frac{1}{36}r^2 \; \overline{{\mathcal G}_5} -\frac{1}{9}r^3\;
\overline{{\mathcal G}_{8}},\\ \nonumber a_4&=&-\frac{1}{180}r^2(-1
+ r + r^2) \; \overline{{\mathcal G}_{12}} -\frac{1}{108}r^3\;
\overline{{\mathcal G}_{14}}\\\label{rf2e} &&-\frac{1}{108}r^3\;
\overline{{\mathcal G}_{16}} +\frac{1}{27}r^4\; \overline{{\mathcal
G}_{23}} \;.
\end{eqnarray}
The $a_n,\;n=5,6,7$ are given in Supplemental Material 1.\cite{supp1}

\subsection{Magnetic moments and susceptibility}\label{sec:R_susc}
To obtain the magnetic moments $\mu_n$ we will adopt a special
method which is available if one knows the moments $t_n$ for all
values of the coupling constants $J_{\mu\nu}$. We replace $H$ by the
one parameter family of Hamiltonians $H_\alpha\equiv
H+\frac{\alpha}{2}\left({\bf S}^2-N r \right)$. Equivalently we can
substitute $J_{\mu\nu}\mapsto J_{\mu\nu}+\alpha$ for all coupling
constants. The magnetic moments then result from differentiating
$\mbox{Tr}(H_\alpha^{n+1})$ w.~r.~t.~$\alpha$ and finally setting
$\alpha=0$:
\begin{eqnarray}\label{r7a}
\left.\frac{\partial}{\partial\alpha}\;\mbox{Tr}\left(H_\alpha^{n+1}\right)\right|_{\alpha=0}&=&
\frac{n+1}{2}\mbox{Tr}\left(H_0^n({\bf S}^2-N r)\right)\\
&=&\frac{(n+1)(2s+1)^N}{2}\left( 3\mu_n - N r t_n\right)\,.\nonumber
\label{r7b}
\end{eqnarray}
We can calculate the left hand side of (\ref{r7a}) if we insert the
results for the moments and consider ``derivatives" ${\mathcal G}'$
of graphs defined in the following way. Let ${\mathcal G}^{(ij)}$
denote the graph ${\mathcal G}$ but with one bond removed,
${\mathcal N}(i,j)\mapsto {\mathcal N}(i,j)-1$. If ${\mathcal
N}(i,j)=0$ then we set ${\mathcal G}^{(ij)}=0$. Further let
$G({\mathcal G})$ and $G({\mathcal G}^{(ij)})$ denote the respective
symmetry groups. Then we define
\begin{equation}\label{r8}
{\mathcal G}' = \sum_{i<j}{\mathcal N}(i,j)\;{\mathcal G}^{(ij)}
\;\frac{|G({\mathcal G}^{(ij)})|}{|G({\mathcal G})|} \;.
\end{equation}
One has, so to speak, to break each bond of the graph and to sum
over all results. Further, one has to introduce factors which
compensate for the possible change of symmetries. For example,
$\,${\includegraphics[width=7mm]{G16}}$ \,'= 6\;
${\includegraphics[width=7mm]{G8}}$\;
+\;${\includegraphics[width=14mm]{G6}}$\;$. It is obvious that the
evaluation of ${\mathcal G}'$ just yields
$\left.\frac{\partial}{\partial\alpha}\;\overline{\mathcal
G}\right|_{\alpha=0}$. Then it is a straightforward task to
calculate the magnetic moments $\mu_0,\ldots,\mu_7$ by using the
above results for the $t_n$. We will display the results for
$\mu_n,\,n=0,1,2,3$ and give the remaining $\mu_n,\,n=4,5,6,7$ in
Supplemental Material 1.\cite{supp1}
\begin{eqnarray}\label{rma}
\mu_0&=&\frac{Nr}{3},\\ \label{rmb} \mu_1&=& \frac{2}{9} r^2
\,\overline{{\mathcal G}_1},\\ \label{rmc} \mu_2&=& \frac{1}{9} r^2
(Nr-1)\,\overline{{\mathcal G}_2}+\frac{4}{27} r^3
\,\overline{{\mathcal G}_3},\\ \nonumber \mu_3 &=& \frac{1}{90} r^2
(8 - (8+5N)r + 12 r^2)\,\overline{{\mathcal G}_5} +\frac{1}{9} r^3
(-1+2r)\,\overline{{\mathcal G}_6}\\ \label{rmd} &+&\frac{2}{9}
r^3(-1+Nr) \,\overline{{\mathcal G}_8} +\frac{4}{27} r^4 \,
\overline{{\mathcal G}_9} +\frac{2}{9} r^4 \, \overline{{\mathcal
G}_{10}} \;.
\end{eqnarray}

The coefficients of the high temperature expansion of $\chi=\beta
\frac{\mbox{\scriptsize Tr}({S}^{(3)2}\exp(-\beta
H))}{\mbox{\scriptsize Tr}(\exp(-\beta H))}$ can be expressed
through the $\mu_n$ and the $t_n$ which occur as coefficients of the
series in the numerator or in the denominator, respectively. The
first four coefficients are given by:
\begin{eqnarray} \nonumber
\chi&=&\sum_{n=1}^\infty c_n\;\beta^n\\ \nonumber
&=&\mu_0\;\beta-\mu_1\;\beta^2 +
\frac{1}{2}(\mu_2-\mu_0\,t_2)\beta^3\\ &+&
\frac{1}{6}(t_3\mu_0+3 t_2 \mu_1-\mu_3)\beta^4 +\ldots. \label{r9}
\end{eqnarray}
Inserting the known values for the $t_n$ and the $\mu_n$ yields the
desired results for the $c_n$. Similarly as in
Sec.~\ref{sec:R_f}, a variety of product rules can be used to simplify
the resulting expressions revealing the extensive character of the
$c_n$.

We will represent the results for the susceptibility's HTE up to
fourth order in the inverse temperature $\beta$. The higher
coefficients $c_n,\,n=5,6,7,8$ are given in Supplemental Material 1.\cite{supp1}

\begin{eqnarray}\label{r10a}
c_1&=&\frac{Nr}{3},\\ \label{r10b} c_2&=& -\frac{2}{9} r^2
\,\overline{{\mathcal G}_1},\\ \label{r10c} c_3&=& -\frac{1}{18} r^2
\,\overline{{\mathcal G}_2}+\frac{2}{27} r^3 \,\overline{{\mathcal
G}_3},\\ \nonumber c_4 &=& \frac{2}{135} r^2 (-1 + r +
r^2)\,\overline{{\mathcal G}_5} +\frac{1}{54} r^3
\,\overline{{\mathcal G}_6}\\ \label{r10d} &+&\frac{1}{27} r^3
\,\overline{{\mathcal G}_8} -\frac{2}{81} r^4 \, \overline{{\mathcal
G}_9} \;.
\end{eqnarray}

\section{Application to frustrated Heisenberg
systems} \label{applic}

To improve the HTE approximation G.~A.~Baker has introduced  Pad\'e
approximants \cite{baker61} (see also
Refs.~\onlinecite{domb_green} and \onlinecite{OHZ06}). These ratios of two
polynomials $[m,n]=P_m(x)/R_n(x)$ of degree $m$ and $n$ provide an
analytic continuation of a function $f(x)$ given by a power series, and,
therefore, they yield a better approximation of the function $f(x)$. As a rule,
approximants with $m=n$ provide best results.
Since we have a power series up to eighth order, we use the corresponding
$[4,4]$ Pad\'e approximant.

\subsection{The Keplerate magnetic molecules
}
\begin{figure}[ht!]
\centering
\includegraphics*[clip,width=55mm]{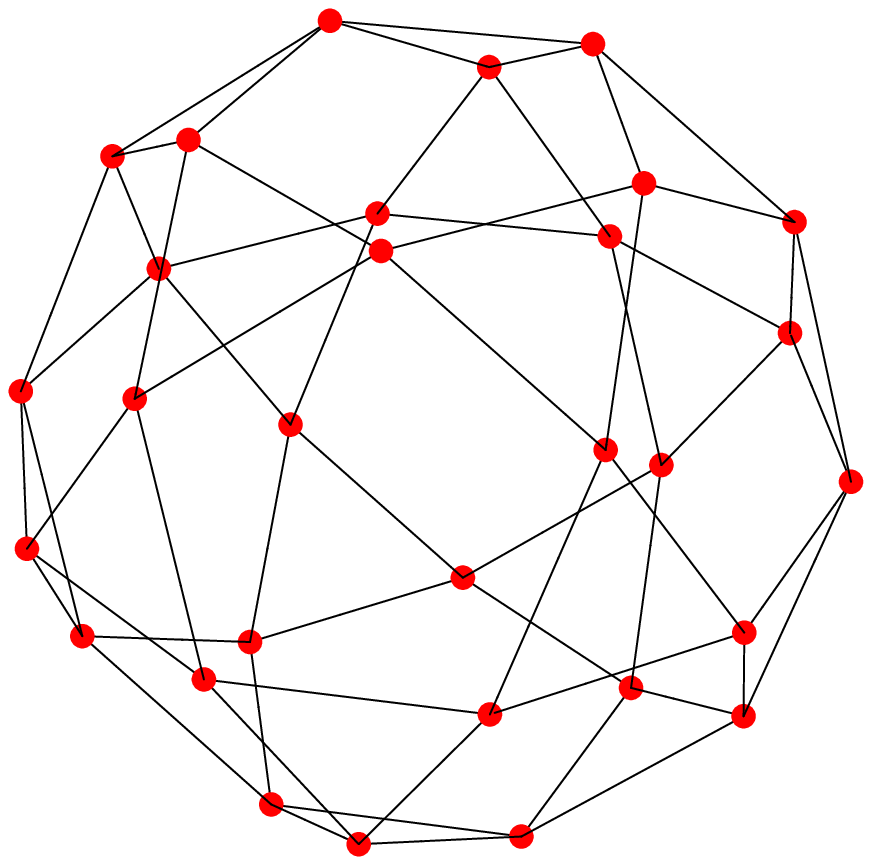}
\caption{The  Archimedean solid icosidodecahedron. In the magnetic molecules
\mofe, \mocr, \mov, \wv\  the magnetic ions occupy the vertices  (red
bullets).
}
\label{icosi}
\end{figure}

In the Keplerate molecules \mofe, \mocr, \mov, and \wv\
the magnetic ions sit on the vertices of an almost perfect
icosidodecahedron\cite{muller2001,olli2005,Cr30,M_W_V30}; see Fig.~\ref{icosi}.
Moreover, the interactions between the magnetic ions are well described by
the Heisenberg model (\ref{model}) with NN interactions.
%===================    figure   =================================
%===================    figure  =================================
These molecules  have attracted much attention from the
experimental \cite{muller2001,olli2005,olli2007,Cr30,fe30_exp_clas_MC,M_W_V30} and
theoretical
side \cite{wir,fe30_spectrum,olli2007,fe30_exp_clas_MC,MC_clas,high_field,finite_T_lanc_a,finite_T_lanc_b}.
One reason is that their frustrated exchange
geometry has much in common
with the kagom\'e lattice, see, e.~g.~, Refs.~\onlinecite{rev1} and \onlinecite{rev2}.
For the \mov\ and \wv\ molecules the spin quantum number is $s=1/2$, which allows us
to calculate low-energy states  exactly by Lanczos exact diagonalization
\cite{wir}. For spin quantum numbers $s>1/2$ that is already impossible,
i.~e.~, for
  \mocr\ ($s=3/2$) and  \mofe\  ($s=5/2$) the low-energy spectrum can be
found only approximately \cite{fe30_spectrum,olli2007}.
To evaluate thermodynamic properties already for $s=1/2$ one has to use
approximations \cite{finite_T_lanc_a,finite_T_lanc_b}. Only at  high magnetic fields
and low temperatures numerical exact results were reported \cite{high_field}.
Very recently a finite-temperature Lanczos approximation has been
used \cite{finite_T_lanc_b} to
describe the magnetic properties
of \wv\ at finite temperatures, and it has been found that  the theoretical results agree
well with the experimental data over a wide temperature range.
However, for frustrated quantum spin systems with $s>1/2$ the calculation
of thermodynamic quantities is even more challenging. Hence our HTE seems
to be useful in particular for $s> 1/2$.
The HTE series for the susceptibility and the specific heat for arbitrary
spin quantum number are given in Eqs.~(\ref{HTE_chi_fe30}) and
(\ref{HTE_C_fe30}) in Appendix~\ref{app_icosi}.

\begin{figure}
\begin{center}
\includegraphics[clip=on,width=80mm,angle=0]{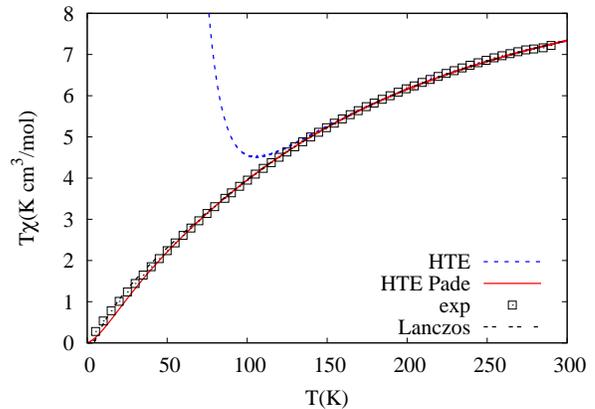}
\end{center}
\caption{Susceptibility times temperature in dependence on the temperature.
The symbols show the experimental data for \mocr\ (Ref.
\onlinecite{M_W_V30}), the black dashed line
represents the finite-temperature Lanczos result \cite{finite_T_lanc_b},
the blue dashed line shows the pure HTE results, and the
red solid line the $[4,4]$ Pad\'e approximant of the HTE series.
For the exchange parameter $J$ and the spectroscopic splitting factor $g$ we
have used the same values as in Refs.~\onlinecite{M_W_V30} and
\onlinecite{finite_T_lanc_b}, namely  $J/k=115$ K and $g=1.95$.
Note that  the Pad\'e approximant and the finite-temperature Lanczos data in
a wide temperature range practically coincide.}
\label{Fe30_compare}
\end{figure}

We focus on the analysis of the HTE data for the susceptibility, since the
high temperature magnetic part of the specific heat often cannot be
accurately separated from the phonon part.
First we compare our $s=1/2$ HTE result for $\chi$
with experimental \cite{M_W_V30} and theoretical \cite{finite_T_lanc_b}
data for \wv.
In Fig.~\ref{Fe30_compare} we show the $T\chi$ vs. $T$ curve as done in
Refs.~\onlinecite{finite_T_lanc_b} and \onlinecite{M_W_V30}.
While the raw HTE data start to deviate from the experimental ones at about
$T=115 $K
 we find an excellent agreement with the experimental results and
the previous theoretical simulations  if we use the $[4,4]$ Pad\'e
approximant.

\begin{figure}[!ht]
\begin{center}
\includegraphics[clip=on,width=75mm,angle=0]{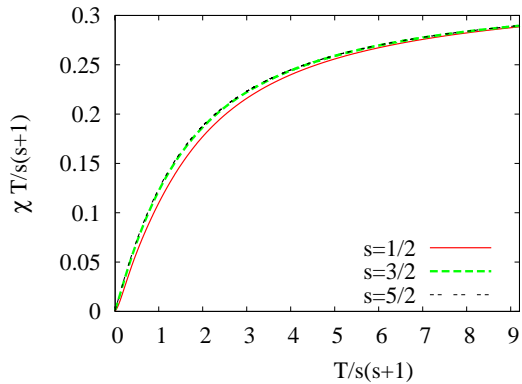}
\end{center}
\caption{Results of the HTE-Pad\'e approximant for the
susceptibility times temperature in dependence on the temperature (arbitrary
units) for spin quantum numbers $s=1/2$, $s=3/2$, $s=5/2$.
}
\label{vgl_Tch_s}
\end{figure}

\begin{figure}[!ht]
\begin{center}
\includegraphics[clip=on,width=75mm,angle=0]{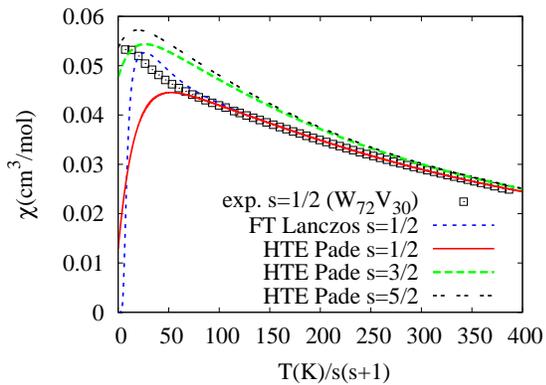}
\end{center}
\caption{Results of the HTE-Pad\'e approximant for the
susceptibility in dependence on the temperature for the Keplerate magnetic
molecule for spin quantum numbers $s=1/2$, $s=3/2$, $s=5/2$ compared with experimental data
for the $s=1/2$ system \wv\ \cite{M_W_V30}
(symbols) and finite-temperature Lanczos results for $s=1/2$
\cite{finite_T_lanc_b} (blue dashed line).
For the exchange parameter $J$ and the spectroscopic splitting factor $g$ we
have used the same values as in Refs.~\onlinecite{M_W_V30} and
\onlinecite{finite_T_lanc_b}, namely  $J/k=115$ K and $g=1.95$.
}
\label{vgl_ch_s}
\end{figure}

Next we compare in Fig.~\ref{vgl_Tch_s} our results for $\chi$ for  the spin quantum
numbers $s=1/2$, $3/2$, $5/2$ relevant for \mov, \wv,  \mocr, and \mofe.  Again we show $T \chi$ vs. $T$, since
such a plot is used in many experimental papers, see, e.~g.~,
Refs.~ \onlinecite{M_W_V30,Cr30,rev1}.
Suggested by Eq.~(\ref{HTE_chi_fe30}) given  in Appendix~\ref{app_icosi}
we use a renormalized temperature $T/s(s+1)$, i.~e.~, we show the dependence
 $T \chi/s(s+1)$ vs. $T/ s(s+1)$ in Fig.~\ref{vgl_Tch_s}.  Obviously the curves for
different
$s$ are very close to
each other. From Eq.~(\ref{HTE_chi_fe30}) it is obvious that with increasing
spin quantum number $s$  in each order of $\beta=1/kT$ the highest-order in
$r=s(s+1)$ yields the dominant contribution, and therefore the plot $T \chi/s(s+1)$
vs. $T/ s(s+1)$ becomes independent of $s$ for larger values of $s$.
However, from both figures \ref{Fe30_compare} and \ref{vgl_Tch_s} the
question arises, whether  the $T \chi/s(s+1)$
vs. $T/ s(s+1)$ plot is appropriate to detect specific features in $\chi$, in
particular at low temperatures. Indeed, the plots in Fig.~\ref{vgl_ch_s}
demonstrate that the characteristic low-temperature maximum in $\chi$ is masked in
the $T \chi/s(s+1)$ vs. $T/ s(s+1)$ plot. The height and the position of the
maximum in $\chi$  clearly depend on $s$. From Fig.~\ref{vgl_ch_s} it is obvious
that its position is shifted to lower values of $T \chi/s(s+1)$ while its
height is increasing with growing $s$.


\subsection{The square-lattice $J_1$-$J_1'$-$J_2$-$J_2'$ model}

Next we consider spin systems  on infinite lattices.
As an example we focus on the frequently discussed square-lattice Heisenberg
magnet with NN couplings $J_1$ and frustrating NNN bonds
$J_2$, the so-called $J_1$-$J_2$ model. This system has
attracted a great deal of interest as a model system to
study quantum phase transitions, see, e.~g.~, the recent publications
\onlinecite{Sir:2006,darradi08,reuther,Richter:2010_ED,richter10_FM,sousa_2010,Johnston}
and references therein. The HTE for the spin-$1/2$
$J_1$-$J_2$ model was presented in Ref.~\onlinecite{Rosner_HTE}.

The interest in this model is also promoted by a number of experimental
investigations on magnetic materials described reasonably well by
the  $J_1$-$J_2$ model.   However, in real materials often one is faced with
deviations from the ideal $J_1$-$J_2$ model. For instance,
in layered vanadium phosphates \cite{rosner2009,rosner2010}
due to low crystal symmetry the bonds along the sides
and the diagonals  of the square can be nonequivalent. Hence,
in a realistic spin
model for these compounds one has to consider two independent NN and two independent NNN
exchange parameters.

\begin{figure}
\begin{center}
\includegraphics[clip=on,width=50mm,angle=0]{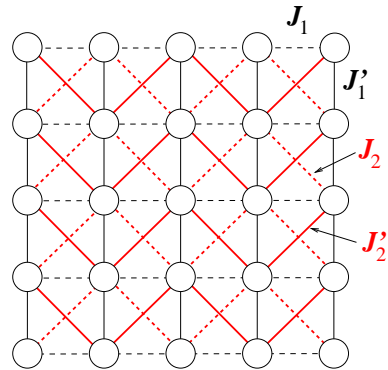}
\end{center}
\caption{Illustration of the exchange paths for the anisotropic frustrated
square-lattice model (\ref{model_square}).
}
\label{fig_model_square}
\end{figure}

Therefore we consider here a generalized  $J_1$-$J_1'$-$J_2$-$J_2'$ model
\begin{eqnarray} \label{model_square}
H&=&J_1\sum_{\langle i,j\rangle} {\bf s}_i \cdot {\bf s}_{j}+
J'_1 \sum_{\langle i,j\rangle'} {\bf s}_i \cdot {\bf s}_{j}
\nonumber \\
&& + J_2\sum_{\langle \langle i,j\rangle\rangle}   {\bf s}_i \cdot {\bf s}_{j}
+ J'_2\sum_{\langle \langle i,j\rangle\rangle'}   {\bf s}_i \cdot {\bf
s}_{j},
\end{eqnarray}
where the NN bonds $J_1$ and $J'_1$ as well as the NNN bonds $J_2$ and
$J'_2$ are arranged on the lattice as shown in Fig.~\ref{fig_model_square}.
This model is more appropriate to provide a realistic description  of frustrated
square-lattice materials such as the layered vanadium phosphates.

\begin{figure}
\begin{center}
\includegraphics[clip=on,width=80mm,angle=0]{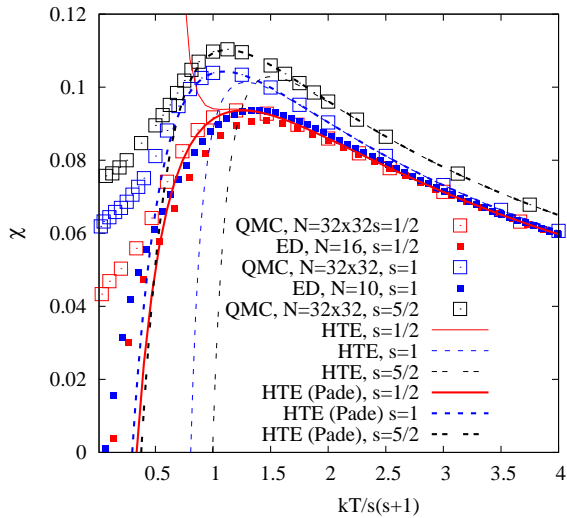}
\end{center}
\caption{Susceptibility $\chi$ of the unfrustrated Heisenberg
antiferromagnet [i.e. $J'_1=J_1=1$,
$J'_2=J_2=0$ in Eq.~(\ref{model_square})] for spin quantum numbers $s=1/2$, $s=1$, and
$s=5/2$. The data of the pure HTE series and the Pad\'e approximant are
compared with corresponding QMC (taken from Ref.~\onlinecite{Johnston}) and ED
results.
}
\label{vgl_QMC}
\end{figure}

Based on our general formulas we get the coefficients of the high-temperature
expansion for the susceptibility and the specific heat for the generalized
model (\ref{model_square}), see the Appendix \ref{app_square} and the
Supplemental Material 2 \cite{supp2}.
These formulas contain also interesting limits of coupled
chain systems \cite{zinke09,volkova,janson2010,nishimoto2011}
obtained by an appropriate choice of the coupling constants.

First, we compare in Fig.~\ref{vgl_QMC} the HTE data for the susceptibility  with accurate QMC data
for the pure square
lattice Heisenberg antiferromagnet  for $s=1/2$ and $s=1$, see, e.~g.~,
Refs.~\onlinecite{troyerPRL1998,troyer1998} and \onlinecite{Johnston}, as well as with numerical exact
data for finite lattices obtained by full exact diagonalization (ED). Again
we use the renormalized temperature $T/s(s+1)$ for the plot; see the
discussion in the previous section.
The comparison with  precise QMC data allows to estimate that
temperature $T_a$ down to which the HTE approximation for $\chi$ is accurate.
We find that the pure HTE in eighth order practically coincides with QMC data
until $T_{a,1}/(J_1s(s+1))=1.2$, $1.4$ and $2.0$ for $s=1/2$, $s=1$,  and $s=5/2$,
respectively. Using the [4,4]
Pad\'e approximant we find $T_{a,2}/(J_1s(s+1))=0.85$, $0.75$, and $0.95$ for $s=1/2$
$s=1$,  and $s=5/2$,
respectively, and it is evident from Fig.~\ref{vgl_QMC} that the
maximum in $\chi$ is described accurately.
Even significantly below $T_{a,2}$ the Pad\'e approximant describes
the QMC data reasonably well.
Moreover, by comparison with ED data we can figure out how good typical ED
results can describe
realistic large systems in two dimensions.
Often, the ED is used as the only method
to discuss the thermodynamics of strongly frustrated 2D quantum spin
systems, see, e.~g.~,
Refs.~\onlinecite{rosner2009} and \onlinecite{shannon04,schmidt07,schmidt07_2}.
The results shown in  Fig.~\ref{vgl_QMC} indicate that for 2D systems already at moderate temperatures and
even for $s=1/2$ (where largest systems are accessible by ED)
significant finite-size effects appear, and that our HTE results for $N \to
\infty$ are better then typical ED results.
A similar finding was reported in Ref.~\onlinecite{rgm2} where ED results
for $\chi$ are compared with data of a Green's function approach
for a spin-$1/2$ frustrated square-lattice ferromagnet.

\begin{figure}
\begin{center}
\includegraphics[clip=on,width=80mm,angle=0]{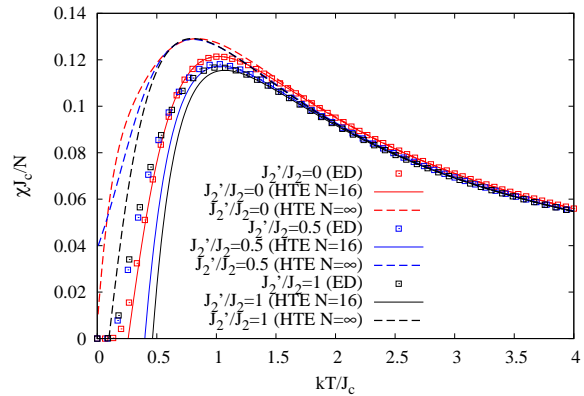}
\end{center}
\caption{Comparison of ED and HTE-Pad\'e results for the susceptibility of the
frustrated model (\ref{model_square}) for spin quantum number $s=1/2$. For
the exchange parameters we follow Ref.~\onlinecite{rosner2009} and choose
$J_1=J_1'=-1/2$ and\\ $(J_2+J_2')/(J_1+J_1')=-2$.
The thermodynamic energy scale is defined as
$J_c=\sqrt{(J_1^2+J_1'^2+J_2^2+J_2'^2)/2}$.
}
\label{vgl_tsirlin}
\end{figure}

We consider now the
generalized  $J_1$-$J_1'$-$J_2$-$J_2'$ model
(\ref{model_square}) relevant for layered vanadium phosphates
\cite{rosner2009,rosner2010}. First, we
mention that, for the symmetric model (i.e. $J_1 = J_1'$, $J_2 = J_2'$), we
give the general formulas for the HTE coefficients for arbitrary $s$ up to
eighth order in  Appendix~\ref{app_square}.
For the
asymmetric model for arbitrary $s$ the formulas become very lengthy for
higher orders. Therefore,  in  Appendix~\ref{app_square} we present
the formulas for arbitrary $s$ only up to
fifth order, and give the remaining sixth to eighth orders in Supplemental Material
2.\cite{supp2}
To illustrate our HTE results
we follow the lines of Ref.~\onlinecite{rosner2009} and discuss the influence of
exchange asymmetry $J_1 \ne J_1'$, $J_2 \ne J_2'$ on the temperature
dependence of the susceptibility, in particular,
on the position and
the height of the maximum in $\chi$.
This issue was discussed Ref.~\onlinecite{rosner2009} based on ED data for
$N=16=4 \times 4$ (see Fig.~9 therein).
We have repeated these ED calculation and compare the ED  results with the
HTE data for $N=16$ and $N \to \infty$ in Fig.~\ref{vgl_tsirlin} and in
Table~\ref{table2}.

\begin{table}
\begin{center}
\caption{Position $T_m=T_{max}/J_c\;$ and height
$\chi_m=\chi_{max}J_c/(N_Ag^2\mu_B^2)$
 of the susceptibility maximum of the model
(\ref{model_square}). In the table we compare the numerical exact ED data
for
$N=16$ (superscript $ED$), the corresponding HTE-Pad\'e data for $N=16$
(superscript $HTE_{16}$) and the HTE-Pad\'e
data for $N=\infty$ (superscript $HTE_{\infty}$).
The thermodynamic energy scale is defined as
$J_c=\sqrt{(J_1^2+J_1'^2+J_2^2+J_2'^2)/2}$. \label{table2}
}
\begin{tabular}{|c|c|c|c|c|c|c|c|c|c|}
\hline
$J_1$ & $J_1'$ & $J_2$ & $J_2'$ & $T^{ED}_m$ & $\chi^{ED}_m$ & $T^{HTE_{16}}_m$ & $\chi^{HTE_{16}}_m $  & $T^{HTE_{\infty}}_m$ & $\chi^{HTE_{\infty}}_m$\\
\hline
$-\frac{1}{2}$ & $-\frac{1}{2}$ & 2 & 0 & 1.02 & 0.1214 & 1.02 & 0.1214 & 0.83 & 0.1290\\
$-\frac{1}{2}$ & $-\frac{1}{2}$ & $\frac{8}{5}$ & $\frac{2}{5}$ & 0.99 & 0.1199 & 1.01 & 0.1197 & 0.81 & 0.1284\\
$-\frac{1}{2}$ & $-\frac{1}{2}$ & $\frac{4}{3}$ & $\frac{2}{3}$ & 1.00 & 0.1181 & 1.04 & 0.1174 & 0.80 & 0.1287\\
$-\frac{1}{2}$ & $-\frac{1}{2}$ & $\frac{8}{7}$ & $\frac{6}{7}$ & 1.02 & 0.1170 & 1.06 & 0.1159 & 0.80 & 0.1291\\
$-\frac{1}{2}$ & $-\frac{1}{2}$ & 1 & 1 & 1.02 & 0.1167 & 1.07 & 0.1155 & 0.80 & 0.1291\\
$-\frac{4}{7}$ & $-\frac{3}{7}$ & 1 & 1 & 1.03 & 0.1165 & 1.08 & 0.1152 & 0.80 & 0.1288\\
$-\frac{2}{3}$ & $-\frac{1}{3}$ & 1 & 1 & 1.05 & 0.1154 & 1.11 & 0.1141 & 0.83 & 0.1276\\
$-\frac{4}{5}$ & $-\frac{1}{5}$ & 1 & 1 & 1.10 & 0.1136 & 1.17 & 0.1118 & 0.89 & 0.1254\\
-1 & 0 & 1 & 1 & 1.15 & 0.1115 & 1.22 & 0.1097 & 0.96 & 0.1225\\
\hline
\end{tabular}
\end{center}
\end{table}

Obviously for $N=16$ the ED  and the corresponding HTE-Pad\'e data for the  maximum
in $\chi$ agree well. But it is also obvious, that the finite-size data for
the maximum do not agree well with data for $N \to \infty$.
The shift of the maximum by
varying the asymmetry (i.~e.~, the difference in $J_1$ and $J_1'$ or/and in
$J_2$ and $J_2'$) discussed Ref.~\onlinecite{rosner2009} is not observed (or is at
least much less pronounced) in the   HTE results  for  $N \to \infty$, cf.
Fig.~\ref{vgl_tsirlin} and
Table~\ref{table2}.
Hence we argue again that the conclusions based on finite-temperature ED data for 2D systems
might be not reliable for large systems.

\subsection{The Heisenberg model on the pyrochlore lattice}

As the last example we consider a 3D frustrated spin system, namely
the Heisenberg model on the pyrochlore lattice. In three dimensions the ED is not
applicable to calculate reasonably well thermodynamic properties. Moreover,
typically there is finite-temperature phase transition which needs special
analysis of the HTE series.
The pyrochlore lattice is highly frustrated and it has attracted much attention over
the last years, see, e.~g.~, Refs.~\onlinecite{moessner01,bramwell01,monopol} and references
therein.
To the best of our knowledge so far no higher-order HTE has been presented.
For the classical limit the thermodynamics was investigated systematically
mainly by classical Monte Carlo (MC)  simulations; see, e.~g.~,
Refs.~\onlinecite{reimers92,moessner99} and \onlinecite{huber01}.  Due to strong
frustration there is no phase transition to an ordered low-temperature phase for the
pyrochlore Heisenberg antiferromagnet.
For the quantum model no precise data are available at lower temperatures.

\begin{figure}
\begin{center}
\includegraphics[clip=on,width=75mm,angle=0]{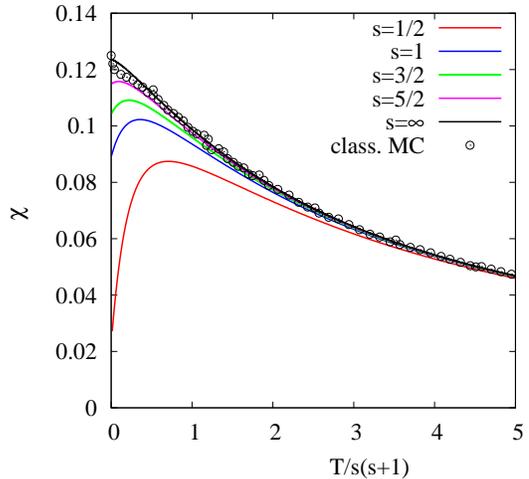}
\end{center}
\caption{HTE Pad\'e approximant for
the susceptibility of pyrochlore Heisenberg
antiferromagnet for $s=1/2$, $1$, $3/2$, $5/2$ and $s \to \infty$.
The MC data  for $s \to \infty$ are taken from
Refs.~\onlinecite{moessner99} and \onlinecite{huber01}.
}
\label{fig_pyro}
\end{figure}

The HTE series for the susceptibility and the specific heat for arbitrary
spin quantum number $s$ are given in Eqs.~(\ref{HTE_chi_pyro}) and
(\ref{HTE_C_pyro}) in Appendix~\ref{app_pyro}.
The plots of the Pad\'e approximants for the Heisenberg antiferromagnet
are shown in  Fig.~\ref{fig_pyro} for various values of $s$.
For the classical model  ($s \to \infty$)  we compare our HTE data with  MC
data calculated in Ref.~\onlinecite{moessner99}, see also
Ref.~\onlinecite{huber01}.
Surprisingly, there is an excellent agreement with the MC data down to
temperatures which are considerably below $|J|/k$. In particular, the fact that
there is no maximum in the $\chi(T)$ curve is observed both in MC and HTE
results.
Lowering the quantum number $s$, i.~e.~, increasing the quantum fluctuations a
low-temperature maximum in $\chi(T)$ emerges. The height $\chi_{m}$ of the maximum
decreases, whereas the position $T_m/s(s+1)$ increases with decreasing of
$s$.\\

\section{Conclusions}
In this paper we provide general expressions for the high-temperature
expansion series  up to eighth order
of free energy, the specific heat,
and the susceptibility for Heisenberg models with arbitrary exchange
patterns $J_{\mu \nu}$ and spin quantum
number $s$. These formulas can be used as a tool to investigate thermodynamic
properties of general Heisenberg systems and thus for the interpretation
of experimental data, especially if other precise methods, such as the quantum
Monte Carlo method or the finite-temperature  density matrix renormalization
group approach, are not applicable.
By comparison with precise quantum Monte Carlo results
for the susceptibility $\chi$ of the unfrustrated
2D Heisenberg antiferromagnet with NN exchange
$J$ with $s=1/2$, $s=1$, ..., $s=5/2$
we find that the HTE results yield
the correct susceptibility at high temperatures down up
to $T/s(s+1) \approx |J|/k$. Using Pad\'e approximants,
the accuracy can be extended to lower temperatures.
In particular, the typical maximum in $\chi$ for the Heisenberg
antiferromagnet can be well described using the HTE of eighth order.

We apply our method to frustrated systems, namely to frustrated Keplerate
magnetic molecules, to a frustrated square-lattice Heisenberg magnet, and to
a pyrochlore Heisenberg magnet.
By comparison with finite-size data
for the unfrustrated as well as the frustrated
square-lattice Heisenberg
model  obtained by full exact diagonalization
we find that the  size of 2D systems accessible by  full exact diagonalization
seems to be too small to get precise data for the susceptibility maximum.
The comparison with Monte-Carlo data for the classical
pyrochlore Heisenberg antiferromagnet yields an excellent agreement down to
low temperatures.  \\

\vspace{1cm}

{\bf Acknowledgement}\\
We thank J. Schnack for providing the data from Ref.~\onlinecite{finite_T_lanc_b} used in
Fig.~\ref{Fe30_compare}.
For the exact diagonalization J. Schulenburg's {\it  spinpack} was used.

\appendix

\section{The high-temperature
expansion for the susceptibility and the specific heat for the Heisenberg
model on the icosidodecahedron with nearest-neighbor exchange interaction
}
\label{app_icosi}
The general formulas for the  susceptibility and the specific heat for the
Heisenberg
model on the icosidodecahedron with the NN exchange constant $J$
up to eighth order read
\begin{widetext}
\begin{eqnarray} \label{HTE_chi_fe30}
&& \chi(\beta)=\frac{N}{J}\sum_{n=1}^\infty c_n(J\beta)^n \; ; \; r=s(s+1)\\
&&c_1=\frac{1}{3}\; r \;\; ; \;\; c_2=-\frac{4}{9}\; r^2 \;\; ; \;\;
c_3=\frac{1}{9}\; r^2(4\; r-1)\;\; ; \;\;
c_4=-\frac{4}{405}\; r^2(3-28\; r+37\; r^2)\nonumber\\
&&c_5=\frac{r^2}{4860}(-45+702\; r-1892\; r^2+1328\; r^3)\nonumber\\
&&c_6=-\frac{r^2}{510\;300}(1728-35\;946\; r+164\;289\; r^2-209\;296\; r^3+99\;776\; r^4)\nonumber\\
&&c_7=\frac{r^2}{2\;041\;200}(-2898+72\;972\; r-467\;127\; r^2+967\;124\; r^3-765\;536\; r^4+259\;008\; r^5)\nonumber\\
&&c_8=-\frac{r^2}{11\;481\;750}(7695-223\;128\; r+1\;769\;382\; r^2-5\;284\;101\;
r^3+6\;231\;056\; r^4 -3\;632\;860\; r^5+745\;760\; r^6)\nonumber
\end{eqnarray}
and
\begin{eqnarray} \label{HTE_C_fe30}
&& C(\beta)=Nk\sum_{n=2}^\infty d_n (J\beta)^n \\
&& d_2=\frac{2}{3}\; r^2 \;\; ; \;\;
d_3=\frac{r^2}{9}(3-4\; r) \;\; ; \;\;
d_4=-\frac{2}{45}\; r^2(-3+23\; r+3\; r^2)\nonumber\\
&&d_5=\frac{r^2}{162}(9-126\; r+116\; r^2+32\; r^3)\nonumber\\
&&d_6=\frac{r^2}{22\;680}(576-11\;142\; r+34\;323\; r^2+5088\; r^3+3952\;
r^4)\nonumber\\
&&d_7=-\frac{r^2}{97\;200}(-1242+29\;556\; r-150\;039\; r^2+100\;736\;
r^3+32\;624\; r^4+25\;472\; r^5)\nonumber\\
&& d_8=-\frac{r^2}{1\;093\;500}(-7695+213\;084\; r-1\;435\;806\; r^2+2\;566\;548\;
r^3+214\;682\; r^4 +473\;600\; r^5+82\;120\; r^6).\nonumber
\end{eqnarray}
\end{widetext}

\section{The high-temperature
expansion for the susceptibility and the specific heat for the square-lattice
$J_1$-$J_1'$-$J_2$-$J_2'$ model
}
\label{app_square}
Here we list the general formulas
for the susceptibility and the specific heat for the
$J_1$-$J_1'$-$J_2$-$J_2'$ model defined in Eq.~(\ref{model_square}).
Since the corresponding formulas become very lengthy in higher orders of the
HTE,
we restrict ourselves here to (i) general formulas for the symmetric model
($J_1=J_1'$ and $J_2=J_2'$) for arbitrary spin quantum number $s$
up to
eighth order and (ii) general formulas for the asymmetric model
($J_1 \ne J_1'$ and $J_2 \ne J_2'$) for arbitrary $s$ up to  fifth order,
only (for
the remaining sixth to eighth order coefficients, see the Supplemental Material
2 \cite{supp2}).
Note that for the symmetric models with $s=1/2$ the HTE coefficients up to
10th  order are given in Ref.~\onlinecite{Rosner_HTE}.

First we give the formulas for the symmetric model:
\begin{widetext}
\begin{eqnarray} \label{HTE_j1j2_chi}
&& \chi(\beta)=N\sum_{n=1}^{\infty}c_n\beta^n \; ; \; r=s(s+1)\\
&& c_1=\frac{r}{3}\quad ; \quad c_2=-\frac{4}{9}r^2(J_1+J_2) \quad ; \quad
c_3=\frac{1}{27}r^2 \Big [ 3J_1^2(-1+4r)+32J_1J_2r+3J_2^2(-1+4r)\Big ]\nonumber\\
&& c_4=-\frac{2}{405}r^2\Big [3J_1^3(2-17r+28r^2)+10J_1^2J_2r(-9+34r)+20J_1J_2^2r(-3+20r)
+3J_2^3(2-17r+28r^2)\Big ]\nonumber\\
&& c_5=\frac{1}{4860}r^2\Big [J_1^4(-45+648r-1808r^2+1712r^3)+120J_1^3J_2r(5-54r+80r^2)\nonumber\\
&&\; \;
+12J_1^2J_2^2r(69-574r+1536r^2)+192J_1J_2^3r(2-27r+68r^2)+J_2^4(-45+648r-1808r^2+1712r^3)\Big ]\nonumber\\
&& c_6=-\frac{1}{127575}r^2\Big [2J_1^5(216-4131r+18339r^2-28710r^3+18100r^4)+21J_1^4J_2r(-279+4801r-14048r^2+12368r^3)\nonumber\\
&&\; \; +14J_1^3J_2^2r(-477+7158r-40044r^2+49864r^3)+35J_1^2J_2^3r(-216+3261r-13504r^2+24240r^3)\nonumber\\
&&\; \;
+350J_1J_2^4r(-9+186r-832r^2+1168r^3)+2J_2^5(216-4131r+18339r^2-28710r^3+18100r^4)\Big ]\nonumber\\
&& c_7=\frac{1}{3061800}r^2\Big[J_1^6(-4347+99738r-623943r^2+1392666r^3-1440944r^4+673152r^5)\nonumber\\
&&\; \; +28J_1^5J_2r(2133-44406r+223896r^2-351328r^3+209664r^4)\nonumber \\
&&\; \;+42J_1^4J_2^2r(1620-35649r+248050r^2-618992r^3+485984r^4)\nonumber\\
&&\; \;+28J_1^3J_2^3r(2061-46134r+389520r^2-1234240r^3+1254912r^4)\nonumber\\
&&\; \;+2J_1^2J_2^4r(36936-789687r+4517328r^2-12128272r^3+15232512r^4)\nonumber\\
&&\; \;+64J_1J_2^5r(432-11097r+75108r^2-181180r^3+166680r^4)\nonumber\\
&&\; \;+J_2^6(-4347+99738r-623943r^2+1392666r^3-1440944r^4+673152r^5)\Big]\nonumber\\
&& c_8=-\frac{1}{91854000}r^2\Big[3J_1^7(20520-536112r+4174761r^2-12734370r^3+18166056r^4\nonumber\\
&&\; \; -13785984r^5+5028608r^6)+8J_1^6J_2r(-108459+2561472r-17865060r^2+42056212r^3\nonumber\\
&&\; \; -43723408r^4+19466016r^5)+8J_1^5J_2^2r(-120123+3081636r-26591049r^2\nonumber\\
&&\; \; +104756322r^3-150488432r^4+82730944r^5)+4J_1^4J_2^3r(-205416+6235740r\nonumber\\
&&\; \; -66481989r^2+275607666r^3-543091408r^4+372472256r^5)+2J_1^3J_2^4r(-364932\nonumber\\
&&\; \; +10618461r-111258000r^2+530418656r^3-1116577664r^4+926078848r^5)\nonumber\\
&&\; \; +20J_1^2J_2^5r(-51678+1313703r-10031442r^2+34441608r^3-66122048r^4\nonumber\\
&&\; \; +61121472r^5)+80J_1J_2^6r(-4347+127044r-1130751r^2+3894570r^3-6116240r^4\nonumber\\
&&\; \; +4059904r^5)+3J_2^7(20520-536112r+4174761r^2-12734370r^3+18166056r^4\nonumber\\
&&\; \; -13785984r^5+5028608r^6)\Big]\nonumber
\end{eqnarray}
and
\begin{eqnarray} \label{HTE_j1j2_C}
&&C(\beta)=Nk\sum_{n=2}^\infty d_n\beta^n \\
&& d_2=\frac{2}{3}(J_1^2+J_2^2)r^2 \quad ; \quad d_3=\frac{1}{3}r^2(J_1^3-8J_1^2J_2r+J_2^3)\nonumber\\
&& d_4=\frac{2}{45}r^2\Big[J_1^4(3-18r+7r^2)-20J_1^3J_2r+50J_1^2J_2^2r(-1+2r)+J_2^4(3-18r+7 r^2)\Big]\nonumber\\
&& d_5=\frac{1}{162}r^2\Big[J_1^5(9-108r+28r^2)-8J_1^4J_2r(9-104r+16r^2)+40J_1^3J_2^2r(-3+5r)\nonumber\\
&& \; \; -4J_1^2J_2^3r(39-274r+256r^2)+J_2^5(9-108r+28r^2)\Big]\nonumber\\
&&d_6=-\frac{1}{22680}r^2\Big[J_1^6(-576+9756r-23739r^2+12160r^3+640r^4)+560J_1^5J_2r(9-135r+16r^2)\nonumber\\
&& \; \; -28J_1^4J_2^2r(-297+4023r-12904r^2+824r^3)+280J_1^3J_2^3r(24-153r+128r^2)\nonumber\\
&& \; \; -70J_1^2J_2^4r(-144+2139r-3872r^2+2400r^3)+J_2^6(-576+9756r-23739r^2+12160r^3+640r^4)\Big]\nonumber\\
&& d_7=\frac{1}{97200}r^2\Big[J_1^7(1242-25920r+116211r^2-58432r^3-1920r^4)+8J_1^6J_2r(-1503
 +28233r\nonumber\\
&& \; \;-99398r^2+31296r^3+3616r^4)+28J_1^5J_2^2r(-639+11076r-31508r^2 +1648r^3)\nonumber\\
&& \;\;+56J_1^4J_2^3r(-279+5943r-42779r^2+59100r^3+64r^4)+42J_1^3J_2^4r(-306+4361r-7144r^2+4000r^3)\nonumber\\
&& \; \; -4J_1^2J_2^5r(5472-110007r+401658r^2-434720r^3 +186560r^4)\nonumber\\
&& \; \;+J_2^7(1242-25920r+116211r^2-58432r^3-1920r^4)\Big]\nonumber\\
&& d_8=\frac{1}{1093500}r^2\Big[J_1^8(7695-185976r+1160352r^2-1811898r^3+889724r^4+20256r^5\nonumber\\
&& \; \; -27512r^6)+8J_1^7J_2r(-10044+212661r-1119348r^2+371668r^3+27120r^4)\nonumber\\
&& \; \; -2J_1^6J_2^2r(57672-1256355r+7596708r^2-19368452r^3+3177376r^4+819808r^5)\nonumber\\
&& \; \; +20J_1^5J_2^3r(-4536+111771r-777870r^2+988176r^3+896r^4)-2J_1^4J_2^4r(44874\nonumber\\
&& \; \; -1369062r+13508715r^2-39201672r^3+31536928r^4+2049424r^5)-20J_1^3J_2^5r(3618\nonumber\\
&& \; \; -69579r+233613r^2-237744r^3+93280r^4)+10J_1^2J_2^6r(-13662+325665r-1954599r^2\nonumber\\
&& \; \; +3232376r^3-2405088r^4+818624r^5)+J_2^8(7695-185976r+1160352r^2-1811898r^3\nonumber\\
&& \; \; +889724r^4+20256r^5-27512r^6)\Big] .\nonumber
\end{eqnarray}
\end{widetext}
\begin{widetext}
Next we give the formulas for the asymmetric model
(up to  fifth order).  For the susceptibility $\chi$ we find
\begin{eqnarray}
&& \chi(\beta)=N\sum_{n=1}^{\infty}c_n\beta^n \; ; \; r=s(s+1)\\
&&c_1=\frac{r}{3} \quad ; \quad c_2=-\frac{2}{9}(J_1'+J_1+J_2'+J_2)r^2;\nonumber\\
&&c_3=\frac{1}{54}r^2\Big[-3(J_1'^2+J_1^2+J_2'^2+J_2^2)+4(J_1'^2+J_1^2+J_2'^2+4J_2'J_2+J_2^2+4J_1(J_2'+J_2)\nonumber\\
&& \; \; +4J_1'(J_1+J_2'+J_2))r\Big];\nonumber\\
&&c_4=\frac{1}{405}r^2\Big[-6(J_1'^3+J_1^3+J_2'^3+J_2^3)+3(7J_1'^3+7J_1^3+10J_1^2(J_2'+J_2)\nonumber\\
&& \; \; +10J_1'^2(J_1+J_2'+J_2)+10J_1(J_2'^2+J_2^2)+(J_2'+J_2)(7J_2'^2+3J_2'J_2+7J_2^2)\nonumber\\
&& \; \; +10J_1'(J_1^2+J_2'^2+J_2^2+J_1(J_2'+J_2)))r-4(J_1'^3+J_1^3+20J_1^2(J_2'+J_2)\nonumber\\
&& \; \; +20J_1'^2(J_1+J_2'+J_2)+20J_1(J_2'^2+3J_2'J_2+J_2^2)+(J_2'+J_2)(J_2'^2+19J_2'J_2+J_2^2)\nonumber\\
&& \; \; +5J_1'(4J_1^2+9J_1(J_2'+J_2)+4(J_2'^2+3J_2'J_2+J_2^2)))r^2\Big];\nonumber\\
&&c_5=\frac{1}{9720}r^2\Big[-45(J_1^4+J_2'^4+J_2^4)+192J_1'^3(J_1+J_2'+J_2)r(1-6r+4r^2)\nonumber\\
&& \; \; +12J_1'r(16 J_1^3+9J_1^2J_2'+9J_1J_2'^2+16J_2'^3+9J_1^2J_2+9J_1J_2^2+16J_2^3-2(48J_1^3\nonumber\\
&& \; \; +87J_1^2(J_2'+J_2)+12(J_2'+J_2)(4J_2'^2+J_2'J_2+4J_2^2)+J_1(87J_2'^2+80J_2'J_2+87J_2^2))r\nonumber\\
&& \; \; +16(J_1+J_2'+J_2)(4J_1^2+4J_2'^2+26J_2'J_2+4J_2^2+17J_1(J_2'+J_2))r^2)-J_1'^4(45+4r(-69\nonumber\\
&& \; \; +4r(11+r)))+12J_1'^2r(30(J_2'^2+J_2^2)-80(J_2'+J_2)^2r+80(2J_2'+J_2)(J_2'+2J_2)r^2\nonumber\\
&& \; \; +10J_1^2(3+8r(-1+2 r))+3J_1(J_2'+J_2)(3+2r(-29+56r)))+4r(48J_2'^3J_2(1-6r\nonumber\\
&& \; \; +4r^2)+48J_2'J_2^3(1-6r+4r^2)+48J_1^3(J_2'+J_2)(1-6r+4r^2)+30J_1^2(3(J_2'^2+J_2^2)\nonumber\\
&& \; \; -8(J_2'+J_2)^2r+8(2J_2'+J_2)(J_2'+2J_2)r^2)+J_1^4(69-4r(11+r))+J_2'^4(69\nonumber\\
&& \; \; -4r(11+r))+J_2^4(69-4r(11+r))+30J_2'^2J_2^2(3+8r(-1+2r))+24J_1(J_2'\nonumber\\
&& \; \; +J_2)(2J_2'^2(1-6r+4r^2)+2J_2^2(1-6r+4r^2)+J_2'J_2(-2+r(-3+52r))))\Big].\nonumber
\end{eqnarray}
For the specific heat $C$ we have
\begin{eqnarray}
&&C(\beta)=Nk\sum_{n=2}^\infty d_n\beta^n \\
&&d_2=\frac{1}{3}r^2(J_1'^2+J_1^2+J_2'^2+J_2^2)\nonumber\\
&&d_3=\frac{1}{6}r^2\Big[J_1'^3+J_1^3+J_2'^3+J_2^3-8J_1'J_1(J_2'+J_2)r\Big]\nonumber\\
&&d_4=\frac{1}{45}r^2\Big[3(J_1'^4+J_1^4+J_2'^4+J_2^4)-2(4J_1'^4+2(2J_1^4+5 J_1^2J_2'^2+2J_2'^4+5(J_1^2+J_2'^2)J_2^2+2J_2^4)\nonumber\\
&& \; \; +5J_1'J_1(J_2'^2+J_2^2+J_1(J_2'+J_2))+5J_1'^2(2J_1^2+J_1(J_2'+J_2)+2(J_2'^2+J_2^2)))r+(-3J_1'^4-3 J_1^4\nonumber\\
&& \; \; -3J_2'^4+20J_2'^2J_2^2-3J_2^4+20J_1^2(J_2'^2+3J_2'J_2+J_2^2)+20J_1'^2(J_1^2+J_2'^2+3J_2'J_2+J_2^2))r^2\Big]\nonumber\\
&&d_5=\frac{1}{324}r^2\Big[9(J_1'^5+J_1^5+J_2'^5+J_2^5)-12(4 J_1'^5+4J_1^5+5J_1^3(J_2'^2+J_2^2)+(J_2'+J_2)(2 J_2'^2-J_2'J_2\nonumber\\
&& \; \; +2J_2^2)^2+5J_1^2(J_2'^3+J_2^3)+5J_1'^2(J_1^3+J_2'^3+J_2^3)+3J_1'J_1(J_2'^3+J_2^3+J_1^2(J_2'+J_2))+J_1'^3(5J_1^2\nonumber\\
&& \; \; +3J_1(J_2'+J_2)+5(J_2'^2+J_2^2)))r+4(-3J_1'^5-3J_1^5-3J_2'^5+10J_2'^3J_2^2+10J_2'^2J_2^3-3J_2^5\nonumber\\
&& \; \; +10J_1^3(J_2'^2+3J_2'J_2+J_2^2)+5J_1^2(J_2'+J_2)(2J_2'^2+J_2'J_2+2J_2^2)+8J_1'J_1(J_2'+J_2)(13(J_1^2+J_2'^2)\nonumber\\
&& \; \; +2 J_2'J_2+13J_2^2)+2J_1'^3(5J_1^2+52J_1(J_2'+J_2)+5(J_2'^2+3J_2'J_2+J_2^2))+5J_1'^2(2J_1^3+(J_2'\nonumber\\
&& \; \; +J_2)(2J_2'^2+J_2'J_2+2J_2^2)))r^2-64J_1'J_1(J_2'+J_2)(J_1'^2+J_1^2+J_2'^2+14J_2'J_2+J_2^2)r^3\Big]\nonumber
\end{eqnarray}
\end{widetext}

\section{The high-temperature
expansion for the susceptibility and the specific heat for the
Heisenberg model on the pyrochlore lattice}
\label{app_pyro}

The general formulas for the  susceptibility and the specific heat for the
Heisenberg
model on the pyrochlore lattice with NN exchange constant $J$
up to eighth order read for the susceptibility as follows:
\begin{widetext}
\begin{eqnarray}  \label{HTE_chi_pyro}
&&\chi(\beta)=\frac{N}{J}\sum_{n=1}^\infty c_n (J \beta )^n\\
&&c_1=\frac{r}{3} \quad ; \quad c_2=-\frac{2r^2}{3} \quad ; \quad c_3=\frac{1}{18}r^2(-3+20r)\nonumber\\
&&c_4=-\frac{1}{135}r^2(6-91r+224r^2)\nonumber\\
&&c_5=\frac{1}{1080}r^2(-15+376r-1816r^2+2544r^3)\nonumber\\
&&c_6=-\frac{1}{14175}r^2(72-2406r+18909r^2-47188r^3+46848r^4)\nonumber\\
&&c_7=\frac{1}{2041200}r^2(-4347+176346r-1901709r^2+7300134r^3-11982944r^4+9482624r^5)\nonumber\\
&&c_8=-\frac{1}{61236000}r^2(61560-2887056r+38320749r^2-202461642r^3+477409712r^4\nonumber\\
&& \; \; -601876480r^5+399408640r^6)\nonumber
\end{eqnarray}
and for the specific heat
\begin{eqnarray}  \label{HTE_C_pyro}
&&C(\beta)=N k \sum_{n=2}^\infty d_n(J\beta)^n\\
&&d_2=r^2 \quad ; \quad d_3=\frac{1}{6}r^2(3-8r) \quad ; \quad d_4=\frac{1}{15}r^2(3-38r+7r^2)\nonumber\\
&&d_5=\frac{1}{36}r^2(3-68r+148r^2+32r^3)\nonumber\\
&&d_6=-\frac{1}{45360}r^2(-1728+53964r-301671r^2+102672r^3+56128r^4)\nonumber\\
&&d_7=-\frac{1}{64800}r^2(-1242+47808r-418437r^2+728520r^3+178240r^4+13312r^5)\nonumber\\
&&d_8=\frac{1}{729000}r^2(7695-345816r+3954204r^2-13638312r^3+5728812r^4+4024640r^5\nonumber\\
&& \; \; +1856680r^6).\nonumber
\end{eqnarray}
\end{widetext}

%\section*{References}

\end{document}